\documentclass{aa}
\usepackage{txfonts}
\usepackage{graphicx}
\usepackage{natbib}
\bibpunct{(}{)}{;}{a}{}{,}

\def\ee{\end{equation}}
\def\be{\begin{equation}}
\newcommand{\beqn}{\begin{eqnarray}}
\newcommand{\eeqn}{\end{eqnarray}}

\begin{document}
\title{Luminosity of a quark star undergoing torsional oscillations\\
and the problem of $\gamma$-ray bursts
}

\author{J. Heyvaerts\inst{3}, S. Bonazzola\inst{1}, M. Bejger\inst{1,2}, P. Haensel\inst{2}}
\institute{
LUTh, UMR 8102 du CNRS, Pl. Jules Janssen, 92195
            Meudon, France
\and
N. Copernicus Astronomical Center, Polish
           Academy of Sciences, Bartycka 18, PL-00-716 Warszawa,
           Poland
\and
Observatoire Astronomique de Strasbourg,
11 rue de l'Universit\'e, 67000 Strasbourg, France}

\offprints{J. Heyvaerts, \\ \email{heyvaerts@newb6.u-strasbg.fr}}
\date{Received 16 November 2007 / Accepted 5 December 2008}
\abstract{}{We discuss whether the winding-up of the magnetic field
by differential rotation in
a new-born quark star can produce a sufficiently-high, energy, emission rate 
of sufficiently long duration to explain long gamma-ray bursts.}
{In the context of
magnetohydrodynamics, we study the torsional oscillations
and energy extraction from a
new-born, hot, differentially-rotating quark star.}
{The new-born compact star is a rapid
rotator that produces a relativistic, leptonic wind.
The star's torsional oscillation modulates this wind emission considerably
when it is odd and of sufficient amplitude,
which is relatively easy to  reach.
Odd oscillations may occur just after the formation of a quark star. 
Other asymmetries can cause similar effects.
The buoyancy of wound-up magnetic fields is inhibited, or its effects are limited, 
by a variety of different mechanisms.
Direct electromagnetic emission by the torsional oscillation
in either an outside vacuum or the
leptonic wind surrounding the compact object is found to
be insignificant. In contrast, the twist given
to the outer magnetic field by an odd torsional oscillation
is generally sufficient to open the star's magnetosphere. 
The Poynting emission of the star in its leptonic environment
is then radiated  from all of its surface and
is enhanced considerably during these open episodes, tapping at the bulk
rotational energy of the star. This results in
intense energy shedding in the first tens of minutes after the collapse
of magnetized quark stars with an
initial poloidal field of order of $10^{14}$ gauss, sufficient to explain
long gamma-ray bursts.
}{}


\titlerunning{Quark star undergoing torsional oscillations}
\authorrunning{J. Heyvaerts et al.}
\maketitle
%%%%%%%%%%%%%%%%%%%%%%%%%%%%%%%%%%%%%%%%%%%%%%%%%%%%%%%%%%
\section{Introduction and motivation}
\label{introduction}
Understanding the physical nature of $\gamma$-ray bursts (GRBs) in a way that is 
consistent with observations of their entire evolution remains a challenging mystery.
A vast literature on the subject exists and a large number of models have been proposed to
explain this phenomenon (for a review see e.g.,
\citealt{ZM04} and \citealt{Mesz06}). In this paper, we limit ourselves to
the case of long GRBs, of duration of between about 10 s and 1000 s.
There are several basic facts to be explained. First, the
release of about 10$^{49} - 10^{51}$ ergs in $\gamma$-rays,
of mean power higher than $10^{48}$ ergs
s$^{-1}$. The violent energy outflow, which eventually transforms into
a GRB, must originate in a compact volume  of
linear size $\sim 10^6~{\rm cm}$, because of the observed
millisecond variability of the GRBs. 
To achieve the observed bulk Lorentz factor $\Gamma =$ 100--1000, an energy outflow of
10$^{51}$ ergs should have a rest-mass load of only 10$^{-5}$--10$^{-6}$ M$_\odot$.
In other words, the baryon wind associated with long GRBs is only 10$^{-8}$--10$^{-6}$ 
M$_\odot$ s$^{-1}$.
Therefore, within the inner engine of GRBs the separation of light from the
matter is realized, producing the  most luminous
electromagnetic explosions in the Universe.

Quark stars are hypothetical stars that consist of deconfined quarks (the
structure of these stars was studied in detail by
\citealt{HZS86} and \citealt{AlcockFO86}). 
They are presumably born in special supernovae, such as SNIc,
from the collapse of very massive Wolf Rayet stars \citep{PH2005}.
The quark star itself forms in a second collapse a few minutes after the new-born 
proto-neutron star simultaneously deleptonizes and spins up. 
Alternatively, a quark star could result from the collapse of
accreting neutron stars in X-ray binaries \citep{ChengDai96}.  The collapse
of stars of an initial mass less than 30 M$_\odot$ is expected to result in the formation 
of a compact object rather than a black hole \citep{Fryer}. Some authors claim that this limit
is in fact 50 M$_\odot$ \citep{Gaensler2005}, which implies that a large fraction 
of the progenitors of SNIc's should  eventually evolve into compact stars.
Among those, a possibly non-negligible fraction could be quark stars \citep{PH2005}.

Bare quark stars differ from the normal,
nucleonic, neutron stars in that the surface of a bare quark star
(a strange star) plays the role of a membrane from which only leptons
and photons  can escape. This property was noted already
in the early papers about quark stars \citep{AlcockFO86,HPA91}.
The quark star surface then effectively separates
baryonic from both leptonic matter and radiation.
The exterior of a quark star is free of baryons, since the latter
would be accreted onto the star
and converted into deconfined quarks, resulting in a release of energy.
The close environment of a newly-born quark star is therefore
expected to be baryon-free.

If quark stars do indeed exist,
they would be prime candidates for emitting
relativistic winds with small baryonic pollution.
The importance of small baryonic pollution in the context of
GRB fireball models was already emphasized
by \citet{HPA91}, when discussing the {\it collision} of quark stars
as an inner engine of short gamma-ray bursts. 
Models of the emission of gamma radiation in a GRB \citep{DeRujula87,ZM04,DarDeRujula04}
require that the central engine emits a bulk relativistic outflow 
with a Lorentz factor $\Gamma$ ranging from 100 to 1000 \citep{Mesz06,Darreview}. 
Such a high Lorentz factor can only be reached if the baryon content of the outflow 
is very low. For example \citet{PaczynskiSuperEdd} has shown that a radiation-driven wind
can reach a Lorentz factor of 100 only if the luminosity injected in the wind
exceeds by a factor 10$^2$ the rest mass energy blown away per second.
\citet{buc06} indicated (using the results of 1D calculation of relativistic winds by
\citealt{Michel69}) that low baryonic pollution is necessary
to obtain high Lorentz factors in a centrifugally-driven magnetized wind. \citealt{dessartetal2007}
reached the same conclusion.
For this reason, the {\it quark star} model is preferable because
the low baryonic pollution of
a quark star's environment ensures that energy can be
deposited cleanly outside the star in the form of
accelerated electron-positron pairs and $\gamma$ ray radiation
without unnecessarily accelerating baryons.
This is why we strongly favour a quark star model 
and consider that the central engine 
of a long GRB may be such a star.
The fact that quark stars could be the source of GRBs was first suggested by
\citet{AlcocketalPhysRevL}.
Our calculations are however not
specific to a quark star, except in Sects. \ref{bulkvisc} and \ref{buoyancy}.
They would also apply to a strongly magnetized neutron star.

The way in which the relativistic wind energy from the quark star is 
eventually released as gamma ray radiation is model-dependent 
and could be due to a range of environments at a distance far 
larger than the light-cylinder radius from the star.
Low baryonic pollution is needed only within a few
light-cylinder radii from the central engine, a few thousand km, 
where the magnetized relativistic wind 
is expected to be accelerated to the required high Lorentz factors. Since the collapse from
proto-neutron star to quark star is slightly delayed, no thick envelope
is expected to be present in the rather small region where the leptonic wind is accelerated.

Newly formed quark stars could be at the origin of GRBs
because of the sudden transformation of hadronic matter into
deconfined quark-matter when a neutron star or a proto-neutron star 
collapses to a quark star. Different ways 
in which such a collapse could be triggered 
have been described by a number of authors
\citep{ChengDai96, ChengDai98a, DaiLu98, ChengDai98b, BombaciDatta00, WangDai00,
OuyedDD02, Lugones2002, OuyedSannino02, Berezhiani02, Berezhiani03, DragoLP04a,
DragoLP04b, PH2005, DragoLP06, DragoLP07, hz07}.

Another promising subclass of models for a GRB central engine is
based on the existence of a rapidly (millisecond period) and
differentially rotating compact star endowed with a
strong (10$^{15}$ to 10$^{17}$ gauss) surface magnetic field, 
or spontaneously developing such a field from differential rotation.
This millisecond magnetar model of the  GRB central engine
was first proposed by \citet{Usov92}, 
who suggested that the GRB energy release
derived its origin from the pulsar activity of a millisecond-period compact object
with a dipole field of 10$^{15}$ gauss.
The idea of a millisecond magnetar has been
revisited and discussed by many authors \citep{Thompson94,ZM01,ZM02}. 
Following \citet{kr98}, a number of authors
proposed differential rotation as a mechanism for strengthening the
toroidal magnetic field in the interior of a newly-born neutron star
\citep{RudermanTK2000} 
or in an accreting neutron star that develops differential rotation as a result of 
an r-mode instability \citep{Spruit99}.
Amplified magnetic fields, of the order of 10$^{17}$ gauss, would then be brought
to the star's surface by buoyancy forces where the energy would
be emitted in the form of bursts and Poynting flux.
\citet{RudermanTK2000} suggested that the emergence of strong fields would 
generate episodic, pulsar activity from the open magnetosphere of the object.
\citet{dl98} considered this model for quark stars.
However, it would be interesting to identify mechanisms that could generate the GRB phenomenon
in objects with fields weaker than 10$^{17}$ gauss. In this paper,
we propose such a mechanism.

A newly-born, compact star is entirely fluid. It supports differential rotation
and internal oscillations of large amplitude
that act on its internal magnetic field. In particular,  
differential rotation in the star
would create a toroidal field from the poloidal component.
This process is often referred to as an 
$\omega$ process, or winding-up of the magnetic field.
The wound-up field and the differential rotation velocity 
constitute an energy reservoir into which electromagnetic emission could tap.
More generally, the star's internal motions may have some effect on the Poynting power radiated.
The main problem of the quark-magnetar model is to explain how the existence of the wound-up
magnetic field could affect the star's emission, either by direct extraction of the associated energy
or by any other means.

Our aim in this paper is to discuss the efficiency of 
the various ways in which a differentially-rotating, 
magnetized, compact star could shed energy in 
its environment. 

We restrict our attention to the case 
of aligned rotators in which  the magnetic dipole axis is parallel to the rotation axis.
The magnetic-field distribution in the star, although it may be locally structured, is 
regarded as smoothly distributed on a larger scale.

We first show
that field winding-up by differential rotation 
is not a steady, but oscillating process (Sect. \ref{torsionaloscill}). 
When much of the available kinetic energy of the differential rotation
has been transformed into toroidal magnetic energy,
the magnetic tension forces react back and reverse the motion.
By means of this mechanism, any initial differential rotation
develops into a torsional, standing, Alfv\'en wave in the star. In the absence of losses,
the star would oscillate, in the rest frame accompanying the average rotation,
like a torsion pendulum (see for instance, \citealt{BVB07}).
In Sect. \ref{torsionaloscill}, we determine the amplitude of the wave and 
in Sect. \ref{bulkvisc} we
discuss its damping in a quark star
as a result of second viscosity. This damping is found to be small.

We then discuss various  mechanisms by which the energy of the torsional oscillation
could emerge from the star. Buoyancy is a possibility. It may however be quenched 
if, while the buoyant matter moves, 
some of the weak reactions between quarks remain frozen, which occurs 
if the matter is in a colour-superconducting state with a large enough gap, and 
the strange quark is sufficiently massive (Sect. \ref{extraction}). Alternatively, buoyancy may
be inhibited by magnetic stratification or, if it develops, it could 
only redistribute flux in the star 
if the latter is magnetized in bulk.

Since the internal magnetic field is time-variable, it could
conceivably act as an antenna and radiate a large-amplitude, electromagnetic
wave in an external vacuum.
We calculate this radiation
in (Sect. \ref{secttruevacuum})
and find that the emitted power is insufficient to produce a GRB.
Radiation in a leptonic medium surrounding the star is shown to
be equally inefficient (Sect. \ref{sectradleptonicwind}).

We next consider the modulation
by differential rotation of the
DC Poynting emission of the fast, aligned magnetic rotator (Sect. \ref{windmodulation}).
We find that an even oscillation, in which the southern hemisphere oscillates
in phase with respect to the northern hemisphere and 
the magnetic structure has a dipolar-type of symmetry
causes a negligible modulation of the energy output.
However, an odd torsional oscillation, in which the southern hemisphere oscillates
in phase opposition with respect to the northern hemisphere,
easily causes the star's magnetosphere to be blown open in a time-dependent way.
An even oscillation acting on a  magnetospheric structure 
that would not be strictly symmetrical
with respect to the equator has a similar effect.
These openings of the magnetosphere 
cause a modulation of the power emitted in the relativistic wind blown by the fast rotator
that is large enough to meet the requirements of
energy and time scale necessary to explain the GRB phenomenon,
even for moderately magnetized stars, with a field of order of a few 10$^{14}$ gauss
(Sects. \ref{opening}--\ref{dampingodd}). 
A collapse that is strictly symmetrical
with respect to the equator would not excite odd oscillations. However, the existence
of a kick received by neutron stars at their birth indicates that a supernova collapse is
in reality not strictly symmetrical. We show that even a weak, odd oscillation is sufficient to
open the star's magnetosphere during several tens of minutes after the collapse.

We use a Gaussian CGS system of units throughout the paper and 
spherical polar coordinates based on the rotation axis, $r$, $\theta$, and $\phi$, 
where $\theta$ is the colatitude.
The corresponding unit vectors are $\vec{e}_r$, $\vec{e}_\theta$, and $\vec{e}_\phi$.

\section{Torsional oscillation in the star}
\label{torsionaloscill}

Differential rotation necessarily induces,
in an highly  magnetized and conducting star, a torsional oscillation. 
Such Alfv\'enic oscillations in fluid, magnetized, compact stars were studied by, e.~g.,
\citet{Bastrukov}, \citet{Rieutord}, and \citet{BVB07}.
We assume that inside the star the poloidal part of the magnetic field is time-independent.
This is reasonable because 
the strange matter is but weakly compressible. 
We also assume that perfect MHD is valid.

\subsection{Magnetic diffusion time scale}
\label{sigmaelectric}

Perfect MHD is a good approximation when the magnetic
diffusion time $\tau_{B}$ is longer than the timescale of the considered phenomenon. 
The time $\tau_B$ depends on the electrical conductivity $\sigma_e$ and 
the gradient lengthscale of the field which we assume to equal the star's radius $R$, 
such that $\tau_B \approx 4 \pi \sigma_e R^2/c^2$.
The electric conductivity is the sum of the 
electronic and quark conductivity, $\sigma_e(e)$ and $\sigma_e(q)$.
The electron fraction in quark matter depends considerably on the physical conditions. 
The quark conductivity was calculated for normal quarks by \citet{Heiselberg}.
Dynamical screening of transverse interactions by the Landau damping of 
the exchanged gluons is important in determining quark mobility.
The result can be expressed, for normal, massless quarks and a strong coupling 
constant supposedly equal to 0.1, as:
\begin{equation}
\sigma_e(q) \approx 5.6 \ 10^{18} \, T_{10}^{-5/3}  \, (n_b/n_0)^{8/9} \, {\mathrm{s}}^{-1} \, ,
\label{electriccondquarks}
\end{equation}
where $n_b$ is the baryonic number density, $n_0$ the nuclear density (0.17 fm$^{-3}$),
and $T_{10} = T/$10$^{10}$$^\circ$K.
If quarks are colour-superconducting in a 
colour-flavour locked (CFL) state, charge neutrality of the quark component alone
is enforced because pairing is maximized when $n_d=n_u=n_s$. This happens when the mass $m_s$
of the strange quark is insufficiently large compared to the gap $\Delta$. 
In the absence of electrons, the CFL colour-superconductor is an electric insulator \citep{Alford}.
Electron suppression 
results in a difference between the chemical potentials of $s$ and $d$ quarks, 
which must remain limited. 
Electrons are suppressed \citep{Rajagopal} only if:
\begin{equation}
\left\vert \frac{m_s c^2}{ 4 {\overline{\mu}}} - \frac{\mu_s - \mu_d}{2} \right\vert < \Delta \, ,
\label{conditionesupprCFL}
\end{equation}
where ${\overline{\mu}}$ is a mean quark chemical potential.
When the mass of the strange quark is so large that
the inequality (\ref{conditionesupprCFL}) is not satisfied, electrons remain present in 
the colour-superconducting quark matter and this matter is then
a conductor with a conductivity which may be far larger than Eq.(\ref{electriccondquarks}).
If quarks are in a two-flavour colour-superconductor (2SC) state, only $d$ and $u$ quarks of
two colours are paired, and electrons are present in quark matter. Then  $\sigma_e(q)$ differs only
by numerical factors from Eq.(\ref{electriccondquarks}) and $\sigma_e(e)$ does not vanish. 
The magnetic diffusion timescale $\tau_B$
associated with the scale length $R$ and the conductivity (\ref{electriccondquarks})
is about 8 $\times$ 10$^{10}$ seconds for $n_b = n_0$ and $T_{10}=1$. 
This is a lower bound to the true magnetic diffusion timescale,
which is longer than this value when the electronic conductivity does not vanish.
This time is long enough for perfect MHD to be valid on the timescale of a burst. 
The exception to this general conclusion is 
when quarks are in a colour CFL-superconductor state in most of the star,
and the $s$ quark is so light that inequality (\ref{conditionesupprCFL}) holds true,
in which case quark matter is an insulator. 
The star's interior is then electrically inactive and none of the phenomena
described in Sects. (\ref{propertiestorsosc}) and 
(\ref{opening}) occurs.
In this paper, we assume that the star's interior is a good electrical conductor, 
which implies that
for the electrical conductivities implied by 
Eq. (\ref{electriccondquarks}) perfect MHD holds true.

\subsection{Period and amplitude of the torsional oscillation}
\label{propertiestorsosc}

\noindent
In perfect MHD, the evolution equations for the velocity
$\vec{V}$ and for the magnetic field $\vec{B}$ inside the
star can be written, in a Galilean rest frame, as:
\begin{equation}
\label{evolV}
\frac{\partial \vec{V}}{\partial t}+ (\vec{V} \cdot \vec{\nabla}) \vec{V}
- \frac{1}{4\pi \rho}\, (\vec{\nabla}\land\vec{B})
\land \vec{B} 
+ \vec{\nabla}(\epsilon + U ) = 0 \ ,
\end{equation}
\be\label{evolB}
\frac{\partial\vec{B}}{\partial t}-\vec{\nabla}\land(\vec{V}\land\vec{B})=0 \ ,
\ee
where $\rho,~\epsilon$, and $U$ are the baryonic mass density,
the enthalpy (the fluid is supposed to be barotropic), and the
gravitational potential, respectively. 
The equations for the toroidal components of {\bf{V}} and {\bf{B}}
can be written in the form~:
\begin{eqnarray}
\frac{\partial V_\phi}{\partial t} &=& \frac{1}{4\pi \rho\, r \sin \theta}  \
\vec{B}_P \cdot \vec{\nabla} \left( r \sin \theta \ B_\phi\right)
\label{evolVphi} \\
\frac{\partial B_\phi}{\partial t} &=&  r \sin \theta \
\vec{B}_P \cdot \vec{\nabla} \left(\frac{V_\phi}{r \sin \theta}\right)
\label{evolBphi}
\end{eqnarray}
We note that for rigid rotation, i.e. for
$V_{\phi}(r, \theta) = \Omega_* r \sin \theta$ with constant
$\Omega_*$, the r. h. s. of equation (\ref{evolBphi})
vanishes, so that no winding-up of the field in this case occurs.
Winding-up results from differential
rotation with depth
or latitude, or both. The Proudman-Taylor theorem does not apply
when non-potential forces, such as the magnetic-tension force,
are exerted onto the fluid and the motions are non-stationary. 
It is however possible that when
non-potential forces are small compared to pressure forces 
and the timescale of the internal motions is long compared to the star rotation period,
these motions organize themselves such that $V_\phi$ becomes
a function of the axial distance $r \sin \theta$ only.
In that case, the field winding is only possible when the poloidal magnetic field has
a component perpendicular to the rotation axis.
The buildup of the toroidal field $B_\phi$ causes a reaction magnetic-tension force
(the terms on the right hand side of Eq.~(\ref{evolVphi})) to grow.
A torsional oscillation then develops.
The system of Eqs. (\ref{evolVphi}) and (\ref{evolBphi})
for the azimuthal velocity and field inside the star has been solved
numerically with appropriate boundary conditions \citep{BVB07}.

We define $R$ to be the star's radius and $R_{10} = R/$(10 km).
The period $P_T$ of the torsional oscillation is esssentially
that of an Alfv\'en wave with a node at both poles, that is, with a  wavelength $2 \pi R$
in a poloidal field $B_P = B_{P14}\,{\mathrm{10}}^{14}$ gauss.
We adopt ${\mathrm{10}}^{14}$ gauss as a reference value since 
fields of order of between ${\mathrm{5\,10}}^{13}$ and 10$^{14}$ gauss
are commonplace in isolated neutron stars \citep{Haberl} and fields
of several 10$^{14}$ gauss are reported to be
typical of anomalous X-ray pulsars and soft gamma-ray bursters \citep{Ziolkowski}.
We assume that the field
is initially rooted deep inside the star. 
This view is supported by simulations of collapse \citep{Obergau2}, which indicate that the
magnetic field after collapse is concentrated in the inner core.
The density of the medium inside the star is of the
order of the mean star density $\rho_* = M/V$,
$M$ being the mass of the star and $V = 4\pi R^3/3$ its volume. 
The period $P_T$ of the torsional oscillation is
that of an Alfv\'en wave of wavelength $2 \pi R$ in a medium of
density $\rho_*$:
\begin{equation}
P_T = \sqrt{\frac{12 \pi^2 M}{B_P^2 R}}
= {4.9\, {\mathrm{s}}} \
\sqrt{\frac{M/M_{\odot}}{B^2_{P14} \ R_{10}}} \ \cdot
\label{estimationPT}
\end{equation}
This result may also be obtained from a linearization of Eqs.~(\ref{evolVphi})--(\ref{evolBphi}).
The magnetic field, instead of pervading all of the star,
could conceivably  be present only in some superficial layer 
where the density is ${\overline{\rho}} < \rho_*$, although the aforementioned simulations do not
support this. In this case, 
$P_T$ would be smaller than given by the expression in Eq.(\ref{estimationPT}) 
by a factor $\sqrt{\, {\overline{\rho}}/\rho_*}$.
The surface density of a quark star is $4 {\cal{B}}/c^2$, where ${\cal{B}}$
is the bag constant. For ${\cal{B}}=$ 40 MeV fm$^{-3}$, this surface density is
2.8~$\times$~10$^{14}$ g cm$^{-3}$, which is about 0.6 times the average density $\rho_*$.
For comparable poloidal fields, the period 
of a superficial torsional wave is only slightly less than 
the period given by Eq.(\ref{estimationPT}) of a global torsional wave.

The wave amplitude is set by the amount of
energy $W_D$ initially stored in the collapse as
kinetic energy of the differential rotation.
We assume $W_D$ to be a fraction $\alpha_D$ of the order of a few percent
of the star's total rotational energy, $W_*$. 
Simulations by
\citet{Obergau2, Obergau1} indicated that $\alpha_D$ does not exceed 10\% after the collapse
to a neutron star. \citet{Burrowsetal2007} found that the value of $\alpha_D$ is less constrained.
We adopt $\alpha_D = 0.1$ as a representative upper limit.
We define $I_*$ to be the moment of inertia of the star and $I_{45} = I_*/$(10$^{45}$ g cm$^2$),
$P_*$ to be the star's average rotation period, and
$P_{*}$(ms) its value in milliseconds. 
\citet{Obergau2, Obergau1} found that when the collapsed core reaches quasi-equilibrium, 
$P_*$ most often ranges in value between 5 and 40 milliseconds.
\citet{Burrowsetal2007} found that 2 milliseconds is a lower bound. 
Since the quark star forms from the hot neutron star after a second collapse, 
its rotation accelerates by a factor $\sim$ 1.5, according to the
moments of inertia calculated by \citet{MichalPawel}. 
Thus, $P_* =$ 3 ms would be a representative value of its spin period.
The rotational energy of the star being $W_* = I_*\Omega_*^2/2$, the
kinetic energy available from the differential rotation
is $W_D = \alpha_D W_*$. 
This is the energy of the torsional oscillation, if it is global.
For $\alpha_D =$ 0.025, $P_*=$ 3 ms, and $I_{45} =$ 1, $W_D \approx$ 6  10$^{49}$ ergs.

The oscillating toroidal field in the star
is at its maximum amplitude $B_T$ when the energy of the torsional oscillation
$W_D$ is entirely in magnetic form, which implies that, for a global oscillation
of a star of volume $V$,
$V B_{T}^2 = 8 \pi W_D$. Thus~:
\begin{equation}
B_T \approx
{\mathrm{10}}^{17} \ \left(
{{{\mathrm{10}} \, \alpha_D \ I_{45} }\over{P_{*}^2({\mathrm{ms}}) \, R_{10}^3}} \right)^{1/2}
\ {\mathrm{gauss}} \cdot
\label{BTinside}
\end{equation}
For $P_*=$ 3 ms, $\alpha_D =$ 0.025, and the other parameters equal to unity,
$B_T \approx$ 1.7$\times$ 10$^{16}$ gauss.
Thus, for a  poloidal magnetic field of order of a few 10$^{14}$ gauss,
the torsional Alfv\'enic oscillation is non-linear.
Equation (\ref{BTinside}) indicates the  typical amplitude of the toroidal field inside the star.
The field distribution  has however
a certain profile with depth and latitude, so that Eq.(\ref{BTinside})
is not a precise estimate of the sub-surface toroidal field.

The matter's angular velocity in the star
is $\Omega_* +  \delta \Omega$, where $\delta \Omega$ varies 
with position  and time and has
a null time average value. Indicating time averaging by brackets, we have
\begin{equation}
W_* + W_D =
{I_* \over 2} \, \Big(\, \Omega_*^2 +  <\delta \Omega^2>\Big)\ \cdot
\end{equation}
From $W_D = \alpha_D W_*$, it is found that the r.m.s. value
of the time-varying angular velocity equals
$\sqrt{\alpha_D} \, \Omega_*$ and
the velocity amplitude of the torsional wave close to the surface
is about a factor of $\sqrt{\alpha_D}$ times the rotational velocity:
\begin{equation}\label{valeurdeltaV}
\delta V_{\phi} = \sqrt{\alpha_D} \, R \Omega_* \, \cdot
\label{amplitudealphaD}
\end{equation}
When magnetic flux is present only in a superficial layer 
of mass $m$, only that part of the kinetic energy of differential rotation
that develops in this layer feeds the energy of the torsional oscillation.
If this energy is distributed proportionally to mass, the estimate given by 
Eq.(\ref{amplitudealphaD}) of the oscillation's amplitude remains valid.
The radial component of the current in the star is then:
\begin{equation}
j_r = {c \over{4 \pi}} \ \vec{e}_r \cdot {\mathrm{rot}} \vec{B} =
{c \over{4 \pi r \sin \theta}} \  
\frac{\partial}{\partial \theta} \Big( \sin \theta B_\phi \Big)\ \cdot
\end{equation}
If this component of the current does not vanish, a DC current
could flow from the star to  the magnetospheric lepton plasma. In Sect. (\ref{opening}),
we consider the consequences of these currents.

\section{Energy-extraction mechanisms}
\label{extraction}

The energy of the torsional oscillation could 
leak out of the star by a number of different mechanisms. 
We consider these mechanisms in turn and discuss whether they could represent
the origin of long GRBs.

\subsection{Viscous damping of the torsional oscillation}
\label{bulkvisc}

The oscillation could be damped  in the star 
by viscous friction or Ohmic dissipation and
then escape by means of heat conduction and thermal radiation from the surface.
The surface of a quark star at temperatures $T\sim$ 10$^{10}\,$$^\circ$K
is an efficient source of $\gamma$ photons and $e^+e^-$ pairs 
(see e.g. \citet{AksenovMU03} and references therein).
However, Ohmic  dissipation is negligible under 
the conditions assumed in Sect. (\ref{sigmaelectric}).
The shear viscosity of normal quark matter has been calculated by \citet{Heiselberg}.
The viscosity of quark matter with unpaired components is of a comparable order of magnitude. 
The Reynolds number for a scale $R$ and a velocity $\delta v_\phi$ 
given by Eq. (\ref{amplitudealphaD})
is found to be of order 10$^{14}$. Thus, shear-viscous dissipation is negligible.

Even superfluid  quark-matter
suffers bulk viscosity \citep{Madsen}.
An Alfv\'en wave, being non-compressive,
is not damped by  bulk viscosity at the linear approximation, but
the Alfv\'enic torsional oscillation is non-linear.
By non-resonant coupling, its magnetic-pressure gradients generate
a compressive oscillation.
Bulk viscosity acting on this
compressive part of the non-linear oscillation causes a damping which can
be calculated by solving the MHD equations perturbatively to second order.
We simplified the calculation of this damping by considering
an homogeneous medium of mass density $\rho_0$
contained in a Cartesian box with unperturbed density $\rho_0$ and magnetic
field $B_0 \, \vec{e}_z$ and an extension $L = \pi/k$ in the $z$-direction.
The solution to first order is the non-compressive standing Alfv\'en wave.
We define $c_{A0}$ and $c_{S0} \gg c_{A0}$ to be the Alfv\'en and the sound
speed in the unperturbed medium respectively, 
$B_T$ the toroidal magnetic amplitude of the wave and
$\zeta$ the coefficient of bulk viscosity. The
second-order solution brings in the following damping time:
\begin{equation}
\tau_{bv} = \frac{32 \rho_0}{\zeta k^2} \, \frac{c_{S0}^4}{c_{A0}^4}
\, \frac{B_0^2}{B_T^2}\ \cdot
\label{taubv}
\end{equation}
The bulk viscosity in quark-matter depends on the finite time
required by quarks to return to the weak-interaction 
equilibrium after the flavour equilibrium is disturbed
by the leptonless strangeness-changing reaction
\begin{equation}
u + d \leftrightarrow s + u
\label{reactionds}
\end{equation}
Any compression of the medium causes such a disturbance,
because the $s$ quark is more massive than the $u$ and $d$ quarks. 
For colour-superconducting quark matter, the existence
of a gap $\Delta$ drastically reduces the reaction rate when $k_BT \ll \Delta$.
The gap energy $\Delta$, which is uncertain, is between 1 and 50 MeV \citep{Madsen2000}.
According to \citet{Madsen}, the bulk viscosity coefficient $\zeta$ of
quark matter experiencing an harmonic density
perturbation of pulsation $\omega$ is:
\begin{equation}
\zeta \approx \frac{\alpha T^2}{\omega^2 + \beta T^4} \ \, {\mathrm{,}} 
\label{zeta}
\end{equation}
where $T$ is the temperature of the medium.
Equation (\ref{zeta}) is valid only when the Fermi energies of the $s$ and $d$
quarks differ by less than $2 \pi k_B T$.
The coefficients $\alpha$ and $\beta$ are:
\begin{eqnarray}
\alpha &=& (4 \pi^2/9) \ K_F \ m_s^4 c^8 \, k_B^2 \, \mu_d^3 \ \, ,
\label{alphaMKS}\\
\beta &=& (\pi^2/9) \ K_F^2 \ k_B^4 \, \mu_d^6 \, h^6 c^6  
\left(1 + {m_s^2 c^4}/({4 \mu_d^2})\right)^2 \cdot
\label{betaMKS}
\end{eqnarray}
In Eqs. (\ref{alphaMKS})--(\ref{betaMKS}), $h$ is the Planck's constant,
$c$ the speed of light,
$k_B$ the Boltzmann's constant, $m_s$ the mass of
the strange quark, and $\mu_d$ the Fermi energy of the $d$ quarks, which
in a 1M$_\odot$ quark star of radius 10 km is 196 MeV.
The mass $m_s$
is expressed in units of 100 MeV by
\begin{equation}
m_s c^2 = m_{s100} \times  {\mathrm{100 \, MeV}} \, \cdot
\label{defms100}
\end{equation}
The rate of the reactions represented by Eq. (\ref{reactionds})
depends on the weak-coupling parameter \citep{Madsen}:
\begin{equation}
K_F = {\mathrm{3.0781}} \, {\mathrm{10}}^{74} \, {\mathrm{g}}^{-8} \, 
{\mathrm{cm}}^{-19} \, {\mathrm{s}}^{15} \cdot
\end{equation}
The time needed to re-establish
the equality of the chemical potentials of the $s$ and $d$ quarks
after a perturbation
(the strangeness equilibration time) is 
$ \tau_{sg} = (\beta T^4)^{-1/2}$.
Writing $T = T_9$ 10$^9$$^\circ$K, the
strangeness equilibration time is:
\begin{equation}
\tau_{sg} =$ 6.9 10$^{-4} T_9^{-2}\, {\mathrm{s}}\, \cdot
\label{tausg}
\end{equation}
This is far shorter than the period of the torsional wave, which means that
$s$ and $d$ quarks always remain close to the equilibrium of the reaction 
given by Eq. (\ref{reactionds})
when the quarks are in a normal state.
For these representative numbers, $(\mu_s - \mu_d)/2 \pi k_B T$
remains small. This legitimates Eq. (\ref{zeta}), in which $\omega^2$ can
also be neglected so that:
\begin{equation}
\zeta = {\mathrm{2.45 \, 10 }}^{29} \
m_{s100}^{\, 4}\ T_9^{-2} \ \, {\mathrm{g\, cm}}^{-1}
{\mathrm{s}}^{-1} \cdot
\end{equation}
If the quarks are in a  colour-superconducting state with a gap $\Delta$, $K_F$ is reduced
by $\exp(-2 \Delta/k_BT)$ \citep{Madsen2000}. For $\Delta =1$ MeV and $T >$ 2.2 10$^9$$^\circ$K, 
the relaxation  time of reactions (\ref{reactionds}) remains less than the period
of the torsional wave. For lower temperatures, it rapidly becomes much longer. 
For example, at 10$^{9 \, \circ}$K with a gap of 1 MeV, 
this time is 8 10$^6$ sec, which is so long 
compared to the period of the wave that
bulk viscosity is quenched.
The damping time $\tau_{bv}$ in a 10$^{14}$ gauss
poloidal field with a toroidal magnetic field amplitude of 10$^{16}$ gauss
is  given by Eq.~(\ref{taubv}). For normal quarks this time is:
\begin{equation}
\tau_{bv} =  {\mathrm{4.8 \ 10}}^{10} \,
 \ T_9^{2}\ m_{s100}^{-4}\ \, 
{\mathrm{s}}\cdot 
\end{equation}
This is much longer than the torsional wave period, owing to the fact
that the compression in this wave is small, so that bulk-viscous damping is negligible.

\subsection{Flux emergence by magnetic buoyancy}
\label{buoyancy}

The internal magnetic field 
can emerge through the surface of the star
and expand into the quasi-vacuum outside as an electromagnetic signal. 
This point of view is
adopted in the models of a number of authors such as \citet{kr98},
\citet{RudermanTK2000}, \citet{Spruit99}, and \citet{dl98}:
amplified magnetic fields, supposedly of order 10$^{17}$ gauss, would be brought
to the surface of a neutron star by buoyancy forces and the energy will
be emitted in the form of bursts and Poynting flux.

We estimate the flux emergence time for a given magnetic field, assuming that nothing
opposes buoyancy.
Since the field in the wound-up flux tubes is
essentially toroidal, these tubes may be regarded as thin circular annuli centred 
on the axis. The rapid rotation of the star inhibits their expansion or contraction
perpendicular to the axis, so that the flux tubes move parallel to it towards the closest pole. Their
length $l$ remains constant. 
Consider a
flux tube of a small cross section $S$ and length $l$, threaded by a field $B$.
Under perfect MHD conditions,
it conserves its magnetic flux and its baryonic content.
For subsonic motions, it also remains
in pressure equilibrium with its environment.
We define $P_{ext}(z)$ to be the material pressure,
$B_{ext}(z)$ to be the magnetic field in this environment
at an altitude $z$ along the polar axis, and $P_{in}(z)$ to be the material pressure
and $B_{in}(z)$ the magnetic field in the tube when it reaches the altitude $z$.
Total pressure equilibrium implies that:
\begin{equation}
P_{ext}(z)  + \frac{B_{ext}^2(z)}{8 \pi} =   P_{in}(z) + \frac{B_{in}^2(z)}{8 \pi} \ \cdot
\label{Ptotsame}
\end{equation}
The pressure $P_c$ in the inner regions of a quark star is about
10$^{35}$ erg cm$^{-3}$. 
Since the magnetic pressure of a field of 10$^{16}$ gauss is far lower,
the difference between the matter densities $\rho_{in}$ and $\rho_{ext}$ in the tube and
in its environment can be calculated perturbatively:
\begin{equation}
\rho_{ext}(z) - \rho_{in}(z)
\approx \frac{3 (B_{in}^2(z)-B_{ext}^2(z)) }{8 \pi c^2}\, \cdot
\end{equation}
where we have used $dP/d\rho = c^2/3$. The difference in magnetic-energy density should 
be added to derive the difference in total energy density.
Denoting by  $(-g_z)$ the component of gravity parallel to the rotation axis, the vertical motion of the flux tube
is described by the equation
\begin{equation}
\left(\rho_{in}(z)+ \frac{B_{in}^2(z)}{8\pi c^2}\right) \ {\ddot{z}} 
=  - g_z \ \frac{B_{ext}^2(z)-B_{in}^2(z)}{4 \pi c^2}\ \cdot
\label{buoyancyequation}
\end{equation}
The existence of the buoyancy instability depends on the distribution of 
the magnetic field in the star. If the field intensity increases with altitude such
that $B_{in}(z)$  is always less than $B_{ext}(z)$ at a higher level, the distribution of 
flux is stable. By contrast, if the flux tube moves in an unmagnetized environment, 
buoyancy can only be inhibited by density differences 
between the magnetized and non-magnetized medium.
These differences may result from a situation of non-equilibrium of weak interactions in the
moving fluid, as described below.
For $B_{ext} =0$, if we assume 
that $B_{in}$ remains almost constant during the motion and
that $g_z \sim 0.5 \ (GM/R^2)$  and $\rho_{in} \approx \rho_* =
M/V$, we find from Eq. (\ref{buoyancyequation}) that the 
buoyancy time $\tau_b$ needed to raise the tube by about a stellar radius is:
\begin{equation} \label{taubuoyancyfull}
\tau_b \approx \sqrt{\frac{6c^2}{G B^2}} =  {\mathrm{2.8 \, 10^{-2}}} \  {\mathrm{s}} \ \ 
\left(\frac{10^{16} \, {\mathrm{G}} }{B_{in}}\right) \, \cdot
\label{taubuoyancy}
\end{equation}
This is the time required to bring an isolated flux tube to the surface when the toroidal field
has reached the value $B_{in}$. 
After the sudden formation of the quark star,
the differential rotation causes this toroidal field to increase in strength as
$B(t)= B_T \sin(t/P_T)$, where $B_T$ is the maximum toroidal field developed
in the torsional oscillation represented by Eq. (\ref{BTinside}). 
Buoyancy starts to be effective only when the growing toroidal field has reached a value such that 
its buoyancy time in Eq. (\ref{taubuoyancy}) calculated for $B(t)$, 
has become shorter than the age $t$ 
of the new-born quark star. For standard stellar parameters and $B_{P14}=1$, 
$P_{*}=$ 3 ms and $\alpha_D =$ 0.025, this occurs at the buoyancy starting time, 
$t_{st} \approx $ 0.5 s.

The ratio of the buoyancy time $\tau_b$ (Eq.~(\ref{taubuoyancy})) to the
period of the torsional oscillation $P_T$ (Eq.~(\ref{estimationPT})) is:
\begin{equation}
\frac{\tau_b}{P_T} =  
\frac{1}{\pi \sqrt{2}} \, \frac{B_P}{B_{in}} \, \left(\frac{GM}{R c^2}\right)^{-1/2} \, \cdot
\label{taubsurPT}
\end{equation}
where $B_P$ is the poloidal field and $B_{in}$ is the total field.
When the wound-up  magnetic field strength is 
close to its maximum value (Eq.(\ref{BTinside})),
$B_{in} \approx$ 100 $B_P$. This implies that 
when nothing opposes buoyancy, 
wound-up fields in isolated flux tubes float to the star's surface in a time $\tau_b$ shorter than
the period $P_T$ of the oscillation.

However, buoyancy motions are reduced or quenched when the ascending
magnetized matter becomes denser, at the same total pressure, 
than matter in its neighbourhood. This may happen 
when the magnetic field pervades the entire volume of the star 
and the field intensity increases with the altitude $z$.
Another effect opposing buoyancy is when
reactions such as Eq. (\ref{reactionds}) or the $\beta$-decay reactions
\begin{equation}
d \leftrightarrow u + e
\label{reactionbeta}
\end{equation}
cannot reach equilibrium in the buoyancy time $\tau_b$.
The relaxation time  $\tau_{sg}$ of
the strangeness-changing reaction (\ref{reactionds}) 
in normal quark matter is about 
7$\times$ 10$^{-6}$ T$_{10}^{-2}$ seconds (Sect. \ref{bulkvisc}).
For colour-superconducting quark-matter with a gap energy $\Delta$, the
time to achieve equilibrium of the same reactions
is lengthened by a factor $\exp(2\Delta/k_B T)$ \citep{Madsen2000}. It may 
become longer than both $\tau_b$ and $P_T$ if the gap $\Delta$ is 
sufficiently large.
Similarly, the relaxation time $\tau_\beta$ of the
quark $\beta$-decay reactions in Eq. (\ref{reactionbeta})
is, for normal quarks,
$\tau_\beta \approx$ 2.7 $\times$ 10$^{-2}$ $T_{10}^{-4}$ seconds \citep{Iwamoto}. 
For colour-superconducting quark-matter, this time is lengthened by a factor 
$\exp(\Delta/k_B T)$.

The temperature is the parameter controlling whether chemical equilibrium
of the reactions in Eqs. (\ref{reactionds}) and (\ref{reactionbeta}) can be achieved
on a given timescale. When a proto-neutron star collapses into a quark star, 
an energy of about 10$^{53}$ ergs is released, which is reflected in the
initial temperature of the new-born object of $\sim$ 3 $\times$ 10$^{11\, \circ}$K. 
The star is then opaque to neutrinos
\citep{Steineretal2001}. It cools by emitting 
thermal $\nu_e$ and ${\overline{\nu}}_e$'s from a neutrino-sphere, 
thermal photons of frequency higher than the
plasma frequency and lepton pairs. According to 
\citet{Usov01}, thermal, photon emission dominates over lepton emission  
at $T >$ 5 $\times$ 10$^{10\, \circ}$K. 
At 10$^{11\, \circ}$K, the photon  emissivity  is 
barely smaller than that of the black-body.
Adding neutrino and antineutrino thermal emission,
the net effective emissivity at this temperature is 
a little less than a factor of two higher than
the black-body emissivity. 
The star then cools to about 3 $\times$ 10$^{10\, \circ}$K in 0.5 seconds.

Does the non-equilibrium of the strangeness-changing reactions or the $\beta$-reactions 
suppress the ascent of magnetized matter to the surface?
If quarks are in a colour-superconducting state with a gap $\Delta$, 
the matter in the buoyant tube 
retains its original strangeness during its ascent if 
$\tau_b < \tau_{sg} \, \exp(2 \Delta/k_B T)$, where $\tau_{sg}$ is the 
relaxation time of the reactions in Eq. (\ref{reactionds})
in normal, quark matter (Eq. (\ref{tausg})).  This condition is satisfied when: 
\begin{equation}
\Delta({\mathrm{MeV}}) >  T_{10}\, (2 \log_{10}T_{10} + 3.6 - \log_{10}B_{16})\, \cdot 
\label{criteriumgaprelevance}
\end{equation}
The time $\tau_b$ is definitely longer than $\tau_{sg}$ at 
3 $\times$ 10$^{10\, \circ}$K, the temperature when buoyancy starts, meaning
that in the absence of a gap, the reactions in Eq. (\ref{reactionds}) remain in equilibrium.
At this temperature and for $B_{16}  = 1$, the inequality 
of Eq. (\ref{criteriumgaprelevance}) holds true 
when $\Delta$ is larger than about 14 MeV, which is plausible because the gap
could  be as high as  50 MeV \citep{Madsen2000}.
For a gap larger than 14 MeV, the reactions in Eq. (\ref{reactionds}) will remain out
of equilibrium during the buoyancy motion.
Similarly, the $\beta$-decay reactions in Eq. (\ref{reactionbeta}) remain frozen during buoyancy motions
if $\tau_b < \tau_{\beta} \exp(\Delta/k_B T)$. This inequality is satisfied without 
the need for a gap when
$T_{10} < B_{16}^{\, 0.25}$, which is not quite the case for $T=$ 3 10$^{10\, \circ}$K and $B_{16} =1$;
this implies that, at this temperature and in the absence of a gap, 
$\beta$-decay reactions remain more or less 
in equilibrium during buoyancy. 
In the presence of a gap, the inequality $\tau_b <\tau_{\beta} \exp(\Delta/k_B T)$ is satisfied 
if the gap $\Delta$ is such that~:
\begin{equation}
\Delta({\mathrm{MeV}}) > 2 \ T_{10}\, (4 \log_{10}T_{10} - \log_{10}B_{16})\, \cdot
\label{criteriumgaprelevancebeta}
\end{equation}
At 3 $\times$ 10$^{10\, \circ}$K and for $B_{16} = 1$,
the inequality (\ref{criteriumgaprelevancebeta}) holds true
when $\Delta$ is larger than 12 MeV.
For this or a larger gap, the reactions in Eq. (\ref{reactionbeta}) keep out
of equilibrium during the buoyancy motion.
It is surprizing that the gap values which freeze the reactions in Eqs. (\ref{reactionds})
and (\ref{reactionbeta}) on the timescale $\tau_b$ are so close. This results from the fact
that  the relaxation time of reactions in Eq. (\ref{reactionds})
lengthens more rapidly with gap energy than that for the $\beta$-decay reactions in
Eq. (\ref{reactionbeta}).

We therefore have two situations. 
The gap is either less than 10 MeV and both reactions in
Eqs. (\ref{reactionds}) and (\ref{reactionbeta})
reach equilibrium on a timescale 
shorter than the buoyancy timescale when buoyancy starts. In this case, chemical
non-equilibrium has no role in limiting buoyancy.
Otherwise, the gap exceeds 14 MeV and both reactions remain frozen
on the buoyancy timescale. We disregard any intermediate situation.

Depressurized, non-equilibrated, quark matter weighs 
more than equilibrated matter at the same pressure
because its energy density is not minimal.
For a gap larger than 14 MeV, buoyancy is quenched 
when the total energy density $\epsilon_{in}$ in the rising flux tube
(including its magnetic energy density) exceeds
the total energy density $\epsilon_{ext}$ in the ambient medium. 
To illustrate this, we consider a flux tube  reaching a region in the star, at an altitude of $z$, 
where its material pressure is less than at the altitude $z_1$ where it started its ascent.
We assume that during its motion the weak reactions remain frozen. 
The difference between the mass density $\rho_{eq}$
of equilibrated matter at this pressure 
and the mass density $\rho_{fr}$ of the frozen matter at the same pressure 
is expressed  by \citet{hz07} as:
\begin{equation}
\rho_{fr} - \rho_{eq} = f_{ab} \, \rho_{eq} \, \cdot
\end{equation}
They calculated $f_{ab}$ for cold quark-matter in which only the
$\beta$-decay reactions (\ref{reactionbeta}) are frozen. 
However we are interested in a situation where the reactions 
in Eqs. (\ref{reactionds}) and (\ref{reactionbeta})
are both frozen. We find that in this case:
\begin{equation}
f_{ab} = 6.4 \, \times\, 10^{-4} \, m_{s100}^{3.7} \, \cdot
\label{faballfrozen}
\end{equation}
where $m_{s100}$ is defined by Eq. (\ref{defms100}).
The field $B_{in}$ of a flux tube that is still buoyant at an altitude $z$
where  the ambient magnetic field is $B_{ext}$ must be such that
\begin{equation}
B_{in}^2 >  B_{ext}^2 + 4\pi c^2 f_{ab} \, \rho_{eq}(P_{in}) \, ,
\label{condibuoy30}
\end{equation}
where $P_{in}$ is the material pressure in the rising flux tube at this altitude. 
Since $f_{ab}\ll 1$,  we can assume  $P_{in}$ to be almost equal to the total external pressure.
For isolated flux tubes moving through
an unmagnetized medium, Eq. (\ref{condibuoy30}) becomes:
\begin{equation}
B_{in}^2 > 4\pi c^2 f_{ab} \, \rho_{ext}\, ,
\end{equation}
where $\rho_{ext}$ is the mass density in the equilibrated unmagnetized 
environment. At the star's surface
$\rho_{ext} \, c^2 = 4{\cal{B}}$, ${\cal{B}}$ being the bag constant. 
We define ${\cal{B}}_{50} = {\cal{B}}/$ (50 MeV fm$^{-3}$).
We assume that $m_{s}=$100 MeV. If quark matter is to be more stable 
than nucleonic matter, ${\cal{B}}_{50}$ should not exceed 1.6 for free quarks confined in the bag, 
and 1.4 for the QCD coupling constant 0.2 (see, e.g., Fig. 8.2 in \citet{Haenseletal2007}).
With $f_{ab}$ given by Eq. (\ref{faballfrozen}),
the minimum value of a field that would be buoyant near the surface, at a level $z$, is:
\begin{equation}
B_{min,\, top} = 4 \, \times\, 10^{16} \ m_{s100}^{1.85}\ 
{\cal{B}}_{50}^{1/2} \ {\mathrm{gauss}} \, \cdot
\label{Bf0allfrozen}
\end{equation}
Deeper inside the star, at a level $z_1$,
the field in this same flux tube had a value $B_{min}$ due to the conservation
of its magnetic flux and
quark number content. From Eq. (\ref{Ptotsame}), $B_{min} = B_{min,\, top} \ 
(P_{ext}(z_1)/P_{ext}(z))^{3/4}$.
The pressure deep inside the star is taken to be
$\rho_{*} c^2/3 \approx Mc^2/4 \pi R^3$.
The pressure close to the star's surface is the bag constant ${\cal{B}}$. 
The minimum buoyant field deep inside the star, $B_{min}$, is then:
\begin{equation}
B_{min} \! = 8.5 \, \times \, 10^{16} \ m_{s100}^{1.85}\ R_{10}^{-9/4} \ {\cal{B}}_{50}^{- 1/ 4}\, 
\left(\frac{M}{M_{\odot}}\right)^{3/4} 
\ {\mathrm{gauss}}\, \cdot
\label{Bf1allfrozen}
\end{equation}
Taking the moment of inertia of the star
to be $I_* = 0.4 \, M R^2$, the ratio of 
the toroidal field $B_T$ generated by the torsional oscillation  
(Eq.~(\ref{BTinside})) to $B_{min}$ is:
\begin{equation}
\frac{B_T}{B_{min}} \approx 0.95\ 
\frac{(\alpha_D/0.1)^{1/ 2}}{P_*({\mathrm{ms}})}
\left(\frac{M_\odot}{M}\right)^{1/4} \
R_{10}^{7/4} \ m_{s100}^{-1.85}\ {\cal{B}}_{50}^{1/4}\, \cdot
\label{criterebuoyantallfrozen}
\end{equation}
Buoyancy is quenched when $B_T/B_{min} < 1$. For $P_*=$ 3 ms, $\alpha_D=$ 0.025, $M=$ 1 M$_\odot$, and
${\cal{B}}_{50}= 4/5$, this occurs (provided that 
the colour-superconductivity gap exceeds 14 MeV) when 
\begin{equation}
m_sc^2 > {\mathrm{38}} \ \, R_{10} \ {\mathrm{MeV}} \, \cdot
\end{equation}
An important question is whether the buoyant flux is entirely expelled out of the star
with the leptonic wind or whether, although the magnetic field partly emerges, it remains rooted 
in the subsurface layers. 

This depends on how rapidly the magnetic field can diffuse 
through quark matter, which itself depends on its electrical conductivity $\sigma_e$
and the gradient lengthscale $l_M$ of the field. Since the magnetic field 
decreases in the flux tube during its ascent, its cross section at the surface cannot be smaller
than when it started. Since rapid rotation prevents radial motions perpendicular to the rotation axis,
the field scale length $l_M$ of buoyant flux tubes cannot diminish.
The fact that the magnetic diffusion timescale is about 10$^{11}$ s 
for conducting quark matter implies 
that during the first few minutes after the formation of the strange star,
the magnetic flux emerging through the star's surface as a result of buoyancy
remains rooted in quark matter at starspots. 
In this case, magnetic activity 
from the torsional oscillations, as described below, continues 
after the first burst of magnetic buoyancy has brought the inner magnetic field closer to the surface.

If matter is magnetized in bulk, buoyancy assumes the form of a convective instability.
When it develops, the more magnetized material is brought to the star's surface,
while the less magnetized material sinks deeper into the star. This results in 
a redistribution of magnetic field in the star, not in a net loss of flux. 
The end result of the field redistribution
should be close to a state of marginal buoyancy instability.
A fraction of the surface magnetic-flux tubes should emerge from the star, baryonic matter
draining down along the field as it emerges. However, since this matter cannot 
diffuse out of the field, 
the emerging magnetic loops remain connected to the subsurface flux.
The magnetized volume experiences little change in this process, so that a substantial part of the
star's volume remains magnetized, if it was initially, and magnetic activity
from the torsional oscillation persists after the flux redistribution.

In the following, we consider cases when the initial field stratification
in the star is stable against buoyancy or when the colour-superconductivity gap is larger
than 14 MeV and the mass of the strange quark sufficiently high
to inhibit the buoyancy of wound-up magnetic fields. Our results also apply to when 
the star was initially magnetized throughout a substantial fraction of its volume and remained
so after a short, first episode of buoyancy.

\subsection{Direct magnetic dipole radiation}

The time-dependent, internal, stellar magnetic field may be a source of electromagnetic emission   
from the star's  environment, whether a vacuum or  a leptonic plasma.
For example, if the new-born star is an oblique rotator \citep{Usov92}, it will emit
electromagnetic radiation due to the rotation
of its magnetic dipole.
We define $\chi$ to be the angle between the magnetic and rotation axes, $B_p$ the polar field,
and $\Omega_*$ the star's rotation rate. The power emitted in vacuo by
the magnetic dipole rotation is \citep{LandauLifchitz}:
\be \label{PoyntPMD}
{\cal{P}}_{MD} 
= \frac{8 \sin^2 \chi}{3 c^3} \, B_p^2 R^6 \Omega_*^4 \ .
\ee
An orthogonal rotator with a polar field of 10$^{15}$ gauss and a rotation period of 3 ms
would radiate a flux of $1.9 \times 10^{48}~{\rm erg~s^{-1}}$, similar to that required
to explain the high-energy photon emission of the gamma-ray burst.
For a polar field of only 10$^{14}$ gauss the emitted power should decline to
$1.9 \times 10^{46}~{\rm erg~s^{-1}}$, which is insufficient to account for a GRB.
We note that this emission taps the rotational energy of the compact star, which is about
$E_{rot} = I_* \Omega_*^2/2 \approx$ 2.2 10$^{51}$ ergs for a rotation period 
of 3 ms. Even at this  high rate,
the radiation of a 10$^{15}$ gauss millisecond magnetar should last for about 10$^3$ seconds.
In the next section, we discuss whether the internal star's torsional oscillation could
somehow act as a substitute for a rotating, magnetic dipole.

\section{Radiation by torsional oscillation}
\label{sectvacuumrad}

The collapse leads to a state of differential rotation in the star,
the angular velocity varying either with depth or with latitude or both.
In an aligned rotator, somewhat analogously to the rotating oblique dipole,
the oscillating internal toroidal magnetic field may act
as an antenna generating a large-scale electromagnetic wave 
in the star's environment at the period $P_T$
of the torsional star's oscillation. 
We calculate in Sect. \ref{secttruevacuum} the power emitted
in a vacuum environment. The radiation in a 
leptonic wind is considered in Sect. \ref{sectradleptonicwind}.

\subsection{Radiation in a vacuum}
\label{secttruevacuum}

To study the electromagnetic emission from the compact star driven by the
torsional oscillation, we begin by calculating the electromagnetic field
in an external vacuum.
Maxwell's equations are solved outside the star under the
boundary conditions
that $B_r$ and the tangential components of the electric field
$\vec{E}$ are continuous at the star's surface.
The matching of the conditions at the star's surface requires neither $B_{\phi}$ nor $B_\theta$
to be continuous, since a
surface current could support a sharp discontinuity in these components.
We denote by a superscript $<$ (or $>$)
quantities relevant to the inside (or the outside) of the star.
The electric field just below the star's surface is given by the law of perfect conductivity:
\begin{equation}
c \, \vec{E}^< +  \ \vec{v}^< \land \vec{B}^< = 0 \, \cdot 
\end{equation}
Since the velocity of the fluid in the star is assumed to be azimuthal only,
the condition that the tangential components
of the electric field are continuous reduces to $E_\phi^> = 0$ and:
\begin{equation}
c \, E_\theta^> = - v_\phi^< \  B_r^< \ \cdot
\label{Ethetacontinu}
\end{equation}
In the presence of a torsional oscillation in the rotating star, Eq.~(\ref{Ethetacontinu})
has both a time-varying and a constant component. 
The latter determines the time-independent, external electric field, 
while the former determines the outside radiation caused by
the torsional oscillation. The boundary condition in Eq.~(\ref{Ethetacontinu})
determines completely the solution in the vacuum outside the star.
We calculate the electromagnetic field radiated out of the star by the internal
torsional oscillation of pulsation $\Omega_T = 2 \pi/P_T$,
assuming axisymmetry. The toroidal field is then the only time-dependent component
of the outer magnetic field.
Omitting for simplicity the superscripts $>$ which refer to the outside region,
the equations for the electromagnetic fields in this region
are Ampere's and Faraday's  equations in a vacuum, given by:
\begin{equation} \label{MaxwellE}
\frac{\partial \vec{E}}{\partial t} - c \vec{\nabla} \land \vec{B}  = 0 \ ,
\end{equation}
\begin{equation} \label{MaxwellB}
\frac{\partial \vec{B}}{\partial t} + c \vec{\nabla} \land \vec{E} = 0 \ \cdot
\end{equation}
The only non-vanishing and time-dependent component
of the magnetic field is the azimuthal one. It is useful to introduce an angular potential
$\mu(r, \theta, t)$ such that
\begin{equation}
B_\phi = \frac{\partial \mu}{\partial \theta} \ \cdot
\label{mu}
\end{equation}
The system (\ref{MaxwellE})-(\ref{MaxwellB})
reduces to an equation for $B_\phi$ alone, which, for harmonic time-dependance in
the form of $\exp(-i\Omega_Tt)$,
translates
into the Helmholtz equation for $\mu$:
\begin{equation}
c^2 \Delta \mu + \Omega_T^2 \, \mu = 0 \ \cdot
\label{Helmholtz}
\end{equation}
The operator $\Delta$ is the ordinary scalar Laplacian.
We expand $\mu(r, \theta, t)$ in spherical axisymmetric harmonics. The solution for each harmonic
component of degree $\ell$ of $\mu$ is, to an arbitrary multiplicative factor:
\begin{equation}
\mu_{\ell}(r, \theta, t) =  \frac{H^{(1)}_{\ell + 1/2}(k_or)}{\sqrt{k_or}} \
\exp(- i \Omega_T \, t) \, P_{\ell}^0(\cos \theta) \ \cdot
\label{solwavel}
\end{equation}
where $k_o$ is given by the dispersion law of 
free-space electromagnetic waves, that is $k_o = \Omega_T/c$.
In Eq.~(\ref{solwavel}), $H^{(1)}_{\ell + 1/2}(x)$ is a semi-integer Hankel function
and $P_{\ell}^0$ is  the Legendre polynomial of order $\ell$.
The value of $\ell=2$ is the lowest value of
$\ell$ pertinent to our problem. In fact Eq.~(\ref{evolBphi})
illustrates that $B^{<}_{\phi} $ inside the star is generated by
$\vec{V}_{\phi} \land \vec{B}_P$ terms. The dipole
components of the poloidal magnetic field $\vec{B}_P^<$ correspond to the lowest value of
$\ell,~(\ell=1)$, since  $B^{<}_r=B^<_p\cos\theta$ and $B^{<}_{\theta}=-B^<_p\sin\theta$, where
$B_p$ is the polar field. Similarly,
the lowest value of $\ell$ for the time-dependent rotation velocity at the star's surface
is $\ell=1$, when $\delta V^\phi = \delta v^\phi_{eq} \, \sin\theta$, $\delta v^\phi_{eq}$
being the velocity of the time-dependent part of the rotation at the equator. This angular variation
in $\delta V^\phi$ corresponds to a constant-amplitude
modulation of the rigid-body rotation-rate at the star's surface:
$\delta \Omega(R, \theta) = \delta \Omega_{eq}$.
Such an oscillation would  be induced by variations in 
the fluid's angular velocity $\Omega$ with depth.
Differential rotation in latitude corresponds to $\ell > 2$.
Since the vector product of $\delta \vec{V}^\phi$ and the poloidal magnetic field
is expanded in spherical harmonics with $\ell\ge 2$, we
now restrict our attention to emission in the $\ell = 2$ mode.
The semi-integer Hankel functions can be expressed as a sum of a finite number of simple terms.
For $\ell=2$, the solution for $\mu$ which behaves as  an outgoing wave at infinity is:
\be
\label{mu2}
\mu\! =\ \! m_{0} R^3\, 
\left(\frac{k_o^2}{3r}+i\frac{k_o}{r^2}-\frac{1}{r^3}\right)
(1-3\cos^2\theta)\, e^{i(k_or-\Omega_Tt)},
\ee
where $m_{0}$ is a complex factor.
The complete solution can be derived
from Eqs.~(\ref{MaxwellE}), (\ref{MaxwellB}), and  (\ref{mu}) and is written in the following form,
where $B_0$ is a complex amplitude:
\begin{equation}
B^{>}_{\phi} = \frac{R^3 B_0}{2} \left(\frac{k_o^2}{3 r}+i\frac{k_o}{r^2}-\frac{1}{r^3}\right)
\, \sin2\theta \, e^{i(k_or-\Omega_Tt)},
\label{Bphi}
\end{equation}
\begin{equation}
E^{>}_{r} = \frac{R^3B_0c}{i\Omega_T}
\left(\frac{k_o^2}{3 r^2}+i\frac{k_o}{r^3}-\frac{1}{r^4}\right)
\, (1-3\cos^2\theta)\, e^{i(k_or-\Omega_Tt)},
\label{Er}
\end{equation}
\begin{equation}
E^{>}_{\theta}  =\frac{R^3B_0c}{2\, i\Omega_T}
\left[\left(\frac{2}{r^4}-\frac{k_o^2}{ r^2}\right)
+ik_o\left(\frac{k_o^2}{3r}-\frac{2}{r^3}\right)\right]
\sin2\theta \, e^{i(k_or-\Omega_Tt)} \cdot
\label{Etheta}
\end{equation}
The relations (\ref{Bphi})--(\ref{Etheta}) solve the system
of Eqs. (\ref{MaxwellE})--(\ref{MaxwellB}). The complex amplitude $B_0$ is determined from the boundary
conditions, such that
the $\theta$ component of the electric field is continuous
at the star's surface (Eq.~(\ref{Ethetacontinu})).
The time-dependent field component $E^{>}_{\theta}$ of Eq. (\ref{Etheta})
must match the corresponding $\ell=2$  time-dependent part of
the field component $E^{<}_{\theta}$ just below the star's surface, which requires that:
\be
\label{EthetaBc}
c \, E_\theta^>\mid_{r=R} \ =  B_p \, \delta v_{\phi \, eq} \ \sin \theta \cos \theta \ ,
\ee
where $B_p$ is the polar field and $\delta v_{\phi\, eq}$ is the time-dependent part of the $\ell=1$
component of the azimutal velocity at the equator.
By matching the two members of Eq.~(\ref{EthetaBc})
we derive the complex wave amplitude $B_0$.
Since the star's radius is far smaller than the wavelength of
the emitted wave, Eq.~(\ref{EthetaBc}) should be evaluated
to the dominant order in the small parameter $k_oR \approx$ 2 $\times$ 10$^{-5}$,
providing:
\be\label{B0vac}
B_0 = \frac{i\, k_oR}{2}  \ B_p\,  \frac{\delta v_{\phi\, eq}}{c} \ \cdot
\ee
The velocity amplitude of the torsional oscillation is given by Eq. (\ref{valeurdeltaV})
and $k_o = \Omega_T/c$. The modulus of $B_0$ is then:
\be
\mid B_0 \mid \, = \, {\mathrm{6.94}} \,\times \, {\mathrm{10}}^{7}
\sqrt{\frac{\alpha_D}{{\mathrm{10}}^{-1}}}
\left( \frac{{\mathrm{10}}^{-3} \, {\mathrm{s}} }{P_*}\right)\!
\left( \frac{ {\mathrm{10}} \, {\mathrm{s}} }{P_T} \right) B_{p14} \, R^2_{10}\ {\mathrm{gauss}}\ . 
\label{B0vacnumeriq}
\ee
The magnetic amplitude of the wave is far smaller than the sub-surface magnetic field because
of the significant impedance mismatch between the star's interior and the outside vacuum.
Denoting by $v_A^<$ the Alfv\'en speed inside the star, the ratio of these impedances is
$v_A^</c \approx 4 \times$ 10$^{-5}\, B_{p14}$. The peculiarities of the spherical wave solution
for $E_\theta$ are also responsible for the smallness of this amplitude,
which is not set by assuming continuity of
$B_\phi$. The correct boundary condition is Eq.~(\ref{EthetaBc}) and
its fulfilment implies that $B_\phi$ is discontinuous
at the star's surface.

It is interesting to evaluate the Poynting power radiated off the star's surface.
If the low-frequency wave emission can
be represented by radiation in vacuo, the solution of Eqs. (\ref{Bphi})--(\ref{Etheta})
provides an upper bound to
the power that may be dissipated in the star's environment and radiated away as X and $\gamma$
photons. The radial component
$\Phi_P^r$ of the Poynting vector associated with the low-frequency radiation is
\be\label{Poynt}
\Phi^r_P= {1 \over 2} \ \left(\frac{c}{4 \, \pi} \, 
{\cal{R}}e \, (E_{\theta}^* B_\phi)\right) \ ,
\ee
where the superscript $*$ designates the complex conjugate and ${\cal{R}}e$ the real part of
a complex number. The power ${\cal{P}}_{vac}$ radiated
as Poynting flux by the torsional oscillation in its supposedly vacuum environment
can be calculated from Eq.~(\ref{Poynt}) by
integrating $\Phi^r_P$ over the star's surface, taking
Eqs.~(\ref{Bphi})--(\ref{Etheta}) into account.
A number of simplifications occur in this calculation, which  finally infers that:
\be \label{PoyntingPTW}
{\cal{P}}_{vac} = \frac{1}{135 \, c^3} \ \Big(\mid \!B_0\!\mid R^3 \Omega_T^2\Big)^2.
\ee
The magnetic amplitude $B_0$ of the wave is given by Eq.~(\ref{B0vac}) and
$\delta v_{\phi\, eq}$ is given by Eq.~(\ref{valeurdeltaV}). We then have:
\be \label{PTvaclitt}
{\cal{P}}_{vac} = \frac{\alpha_D}{540 \, c^7} \ B_p^2 \, \Omega_T^6 \, \Omega_*^2 \, R^{10}.
\ee
Numerically, the power ${\cal{P}}_{vac}$  amounts to:
\be \label{powervacnum}
{\cal{P}}_{vac} \! = \! {\mathrm{2.06}} \times {\mathrm{10}}^{17} \, B^2_{p14} \, R_{10}^{10}
\left(\frac{\alpha_D}{{\mathrm{10}}^{-1}}\right)
\left( \frac{{\mathrm{10}}^{-3} \, {\mathrm{s}} }{P_*}\right)^2 \!
\left( \frac{ {\mathrm{10}} \, {\mathrm{s}} }{P_T} \right)^6 \, {\mathrm{erg}} \, {\mathrm{s}}^{-1} \cdot
\ee
A glance at Eqs.~(\ref{PoyntPMD}) and (\ref{PTvaclitt})
indicates that much less energy is radiated away in an outside vacuum
by the torsional oscillation than by an oblique, rotating, magnetic dipole.
The Poynting power radiated by  the torsional wave is smaller than
the emission of a rotating dipole for several reasons.
First, the radiation
is quadrupolar instead of dipolar.
Then, the field $\mid \!B_0\! \mid$ is only of the order 10$^{8}$ gauss for the
values of parameters adopted as representative
(Eq.~(\ref{B0vacnumeriq})), much less than the polar field of
an ordinary pulsar or magnetar. Finally, the
period $P_T$ is of the order of a few seconds, much longer than the
rotation period of the new-born compact star.
As a result, the power in Eq. (\ref{powervacnum})
falls short by many orders of magnitude of
the observed power of $\gamma$ radiation in a long GRB~!

\subsection{Radiation in a leptonic wind}
\label{sectradleptonicwind}

Could the presence of a circumstellar, leptonic plasma drastically change
the power radiated by the torsional oscillation?
This plasma originates in charges, electrons, and positrons that have passed
the bag of the quark star.
We note that at a distance from the star of larger than the light-cylinder radius,
the plasma cannot be in rigid corotation, but must flow
outward \citep{GoldJulian}. Since the wavelength associated
with the frequency of the torsional oscillation is much larger than the light-cylinder radius
of the rapidly spinning star, the wave emitted by this oscillation
propagates into the wind driven by the rapid global rotation.
We have to determine the amount of energy of the torsional
oscillation radiated per second in these conditions.
We assume that the wind has already reached its terminal velocity
at the surface of a sphere of radius comparable to that of the light-cylinder, $c/\Omega_*$.
Since the ratio of the torsional oscillation period to
the spin rotation period is large, the torsional oscillation
appears, on both of the scales of the spin period and the light-cylinder radius,
as a quasistatic perturbation. Its effect is not only to emit a signal 
that assumes the character of
a wave at distances larger than $c/\Omega_T$, but it also
modulates the wind in which it propagates as a result
of the variations imposed on the conditions of its lauching.
These modulational effects are distinct from emission of low-frequency radiation in a given wind.
Much of the action causing wind modulation occurs below or close to the light-cylinder and
will be discussed in Sect. \ref{windmodulation}.

We now calculate the emission by the torsional oscillation
in an expanding, possibly resistive, leptonic wind.
The conductivity $\sigma$ of the medium is assumed to be real.
This is because the wave would be highly non-linear if
the gyrofrequency $\omega_{B}$ of leptons in the wave's magnetic field
was much larger than the wave frequency.
As a result, the effect of the plasma current
on the real part of the index of refraction would become negligible and the wave
would force its way non-linearly through the leptonic environment
\citep{Estelle75,Salvati78}. We may then restrict our consideration
to resistive effects.
The modulus of the wind speed $\vec{w}$ is assumed to be constant, both in time and space, and
oriented radially outwards: $\vec{w} = w \, \vec{e}_r$. We assume the wind to be ultra-relativistic
and to have a velocity equal to the speed of light.
When taking the wind to be radial, we assume that corotation is lost at distances of the order
or larger than $c/\Omega_T$.
The background, magnetic field in the
wind is severely wound up by the star's rotation. At distances much larger
than the light-cylinder radius, its azimuthal component dominates over the poloidal
component and declines proportionally to $1/r$. 
We thus neglect the poloidal field component
and assume the unperturbed magnetic field $\vec{B}_0$ to be azimuthal, so that:
\begin{equation}
\vec{B}_0 = {\hat{B}}_0 \, \frac{R}{r} \, f(\theta) \, \vec{e}_\phi \ \cdot
\label{Bazimutal}
\end{equation}
Specifically, we consider
$f(\theta) = \cos \theta/(\mid \cos \theta \mid \, \sin \theta)$.
The electric current associated with Eq. (\ref{Bazimutal}) is radial.
For the adopted angular profile 
it reduces to zero almost everywhere, except at both the polar axis
and the equatorial plane. \citet{HN2003} demonstrated that the magnetic field
in a perfect MHD wind asymptotically becomes potential almost everywhere, the current being confined to
boundary layers about the polar axis and at surfaces where the poloidal polarity reverses. 
Our choice 
of $f(\theta)$ complies  with this. The singularity at the polar axis
represents the  current carried by a jet, while the change in the direction of the field at
the crossing of the equator is caused by the change in polarity of the poloidal field at
$\theta = \pi/2$. Ohm's law infers that:
\begin{equation}
\vec{j} = \sigma  \Big(\vec{E} 
+ \frac{1}{c} \, \vec{w} \land \vec{B}\Big) + \rho_e \vec{w} \ \cdot
\label{Ohmaveccharge}
\end{equation}
By considering the divergence of Eq. (\ref{Ohmaveccharge})
and solving the resulting differential equation for the charge density $\rho_e$,
it is found that the latter
vanishes in the unperturbed wind. The unperturbed electric field then vanishes too.
We define $\vec{E}$, $\vec{B}$, $\vec{j}$, $\rho_e$, and $\vec{u}$ to be the perturbations of
the electric and magnetic field, current density, 
charge density, and leptonic fluid velocity, respectively.
It is sufficient to describe the lepton's dynamics at the inertia-less (also called force-free)
approximation. The magnetic-field perturbation is toroidal in the considered geometry
and the electric force $\rho_e \vec{E}$ is a negligible second-order term.
The perturbed Maxwell equations, Ohm's law, and the dynamical equation in the 
zero-inertia limit can be written
(using the compact notation $\partial_t$ for time derivatives) as:
\begin{equation}
c \vec{\nabla} \land \vec{E} = - \partial_t \vec{B} \ ,
\label{Faradpertwind}
\end{equation}
\begin{equation}
c \vec{\nabla} \land \vec{B} = 4\pi \vec{j} + \partial_t \vec{E} \ ,
\label{Amperepertwind}
\end{equation}
\begin{equation}
\vec{\nabla} \cdot \vec{B} = 0 \ ,
\label{solenopertwind}
\end{equation}
\begin{equation}
\vec{\nabla} \cdot \vec{E} = 4 \pi \rho_e \ ,
\label{Poissonpertwind}
\end{equation}
\begin{equation}
c \vec{j} = \sigma  \Big(c \vec{E} + \vec{w} \land \vec{B} + 
\vec{u} \land \vec{B}_0 \Big) + c \rho_e \vec{w} \ , 
\label{Ohmpertwind}
\end{equation}
\begin{equation}
\vec{j} \land \vec{B}_0  =  0 \ \cdot
\label{forcefreepertwind}
\end{equation}
The toroidal  components of Eqs.~(\ref{Faradpertwind})  and (\ref{Ohmpertwind}) 
imply that the electric field is only poloidal. 
The other two components of Eq. (\ref{Ohmpertwind})
infer the fluid velocity once the solution for the other unknowns has been found.
The current $\vec{j}$ is eliminated by taking the 
vector product of Eq. (\ref{Amperepertwind}) and $\vec{B}_0$.
As a result the conductivity is eliminated from
the equations describing the perturbation.
With Eq.~(\ref{Faradpertwind}), this infers an equation
for the magnetic perturbation, which is expressed most accurately 
in terms of the angular potential
$\mu$ of this perturbation (Eq.~(\ref{mu})). 
For harmonic time-dependence $\exp(-i\Omega_Tt)$,
it is found that $\mu$
satisfies the Helmholtz equation (Eq.~(\ref{Helmholtz})).
Regardless of the conductivity, the perturbations propagate in this geometry as
electromagnetic waves, provided the inertia-less limit is considered.
The complete solution is identical to the vacuum result
(Eqs.~(\ref{Bphi})--(\ref{Etheta})) as is of course the Poynting flux at the star's
surface and the emitted Poynting power (Eq.~(\ref{PTvaclitt})).

\section{Modulation by the torsional oscillation of the energy emitted in the rotator's wind}
\label{windmodulation}

A rapidly-spinning aligned rotator emits
a wind carrying power in electromagnetic,  potential, thermal  and kinetic energy form.
The contributions of these different forms of energy depend on the distance
to the star. Some forms of energy may dissipate en route or at terminal shocks, producing
observable X and $\gamma$ radiation. Close to the compact star, much of this flux
is in Poynting form because the kinetic energy remains low while
the wind has not yet been effectively accelerated.
The thermal and gravitational energy fluxes often constitute
but a little part of the total energy flux.
The energy output of the star in its wind environment then enters
the latter as DC Poynting flux,
the radial component of which is given in terms of the field components just above
the star's surface by:
\begin{equation}
\Phi_{DC} = \frac{c}{4 \pi} \, E_\theta^> B_\phi^> \ \cdot
\end{equation}
The boundary condition at the star's surface implies that
$E_\theta^> = E_\theta^<$.
We define $v_\phi$  to be the subsurface fluid velocity and  $B_r$ the radial field component,
which is continuous across the star's surface. We then find that:
\begin{equation}
\Phi_{DC} =   - \, \frac{1}{4 \pi} \ \, v_\phi(R) \ B_r(R) \  B_\phi^>(R) \ \cdot
\label{phiDC}
\end{equation}
The value of $B_\phi^>$ at the base of a relativistic wind is given approximately 
by Eq.~(\ref{BphisurBr}) below.
This can be sketchily explained as follows:
due to the rotation of the star and the effect of flux freezing,
a toroidal field is generated on open field lines from the poloidal field.
The knowledge that the foot point of a field line is rotating is propagated along this line
at a finite velocity $v_{prop}$, by means of convective transport and 
propagation as an Alfv\'enic signal.
The field line thus curves away from
the sense of rotation at an angle of $\xi$ to the radial 
direction, such that $\tan \xi = \Omega_* R\sin \theta /v_{prop}$.
Since the wind is relativistic and the Alfv\'en speed in the
tenuous external plasma is close to the speed of light, $v_{prop}\sim c$ and
the toroidal field just above the star's surface is:
\begin{equation}
B^>_\phi(R) \approx - \, \frac{R\, \Omega_*\sin \theta}{c} \ B_r(R) \ \cdot
\label{BphisurBr}
\end{equation}
A more precise
justification of the approximate relation in Eq. (\ref{BphisurBr}) is
omitted for conciseness.
Then, from Eq.~(\ref{phiDC}):
\begin{equation}
\Phi_{DC} \approx \frac{v_\phi^2 B_r^2}{4 \pi c} \ \cdot
\label{PoyntingDCgeneral}
\end{equation}
This flux is only emitted from the polar caps, the
regions on the stellar surface connected to open, field lines.
The magnetosphere is closed, where the apex of the local field line is at a distance $D$
smaller than the light-cylinder radius $c/\Omega_*$. When field lines are
dipolar, the polar caps extend to a colatitude $\theta_{pc}$,
which we refer to as classical, and described by:
\begin{equation}
\sin^2\! \theta_{pc} = \frac{\Omega_* R}{c} \ \cdot
\label{thetapc}
\end{equation}
Under certain conditions however,
the magnetospheric field may depart considerably from dipolarity (Sect.~\ref{opening}).
When the flux is distributed on the star
as a dipolar field, the radial, field component varies with $\theta$ as
$B_r = B_p \cos \theta$, $B_p$ being the field at the pole.
By considering $v_\phi(\theta)$ to be the solid body rotation velocity at the angular speed $\Omega_*$,
we obtain, by integrating
over the colatitudes corresponding to the two polar caps, the DC Poynting power emitted by the star
under these conditions:
\be\label{PoyntDC}
{\cal{P}}_{sol} =  \frac{B_p^2 R^6 \Omega_*^4}{4 c^3} \ \cdot
\ee
The power represented by Eq. (\ref{PoyntDC}) is comparable to the power emitted
by an oblique, rotating dipole (Eq.~(\ref{PoyntPMD})).
This is a classical result (see for example \citet{Michelbook}).
With a rotation period of 3 ms,
the power ${\cal{P}}_{sol} \approx$ 1.78 $\times$ 10$^{45} B_{p14}^2$ erg s$^{-1}$. 
This is insufficient
to match the high luminosity of a GRB unless fields in excess of 10$^{15}$ 
gauss are involved \citep{Usov92}.
Could differential rotation drastically change this result~?

\subsection{Quasistatic modulation of the wind}
\label{sectevenmodu}

We assume that the star experiences a torsional oscillation. The velocity  $v_\phi(\theta)$
then differs from the solid-body rotation velocity and varies with time. Because
the period of the oscillation is much longer than the mean spin period,
this causes a quasistatic change in both the structure of the magnetosphere
and the polar cap angle. It even causes a change, which we neglect,
in the shape of the light cylinder.
The structure of the magnetosphere and the energy output of the
wind adjust to equilibrium
values corresponding to the instantaneous velocity profile on the star's surface.

This profile may be even with respect to the equator, or odd, or a mixture of both.
An even oscillation is one in which the azimuthal velocity perturbation is
symmetric with respect to the equator, i.e. where the time-varying azimuthal velocity
is in phase at two points positioned symmetrically with respect to the equator. 
An odd oscillation is one in which
it is antisymmetrical, i.e. where the velocity is in phase-opposition
at two points symmetric with respect to the equator.
It may appear that nature should provide only even profiles,
by a principle, or rather a postulate, of north-south symmetry.
However, this symmetry is broken in the case of the collapse of
supernovae. It is indeed well known that new-born neutron stars
receive a kick, that is, a net thrust from the collapse.
This  would be  forbidden by the principle of north-south symmetry.
Therefore, it cannot be excluded that, similarly, odd modes of torsional oscillation are present in the
initial excitation of a new-born compact star. The amplitude of these modes
is expected to be small, but we show below that it need not
be large to produce important effects. The kick received by a new-born neutron
star produces a velocity of the order of 200 - 500 km s$^{-1}$. 
This corresponds to an asymmetry in the momentum emission of the order of
1M$_\odot \times$ 300 km s$^{-1}$.
The supernova explosion emits a momentum per steradian of about
10 M$_\odot \times$ 30 000 km s$^{-1}/$ 4 $\pi$. The asymmetry in the momentum
emission appears to be a fraction of between a few 10$^{-4}$ and a few 10$^{-3}$ of the total.
A similar fraction of the total star's rotational energy may appear
after the collapse in the form of odd differential rotation.

\subsection{Modulation of the wind by an even oscillation}
\label{modupaire}

An even torsional oscillation has but little effect on the structure of the magnetosphere and wind
because the two footpoints of a closed field line follow the same motion exactly 
if, as assumed in this subsection,
the poloidal field lines are 
strictly symmetric with respect to the equator. Otherwise,  
these field lines would undergo a twist in the presence of 
an even torsional oscillation, because their footpoints would not be at exactly 
opposite latitudes, and would be carried in the azimuthal direction at different 
angular velocities.
It is difficult to realistically anticipate the degree
af asymmetry in poloidal field lines. It could vary from very little 
to a complete absence of symmetry. 
The degree of asymmetry necessary for the magnetosphere to open
will be estimated in Sect. (\ref{opening}).
Strictly symmetric field lines whose footpoints are moved by an even torsional oscillation
are, however, not twisted and, as a result, no poloidal electric current is driven
in the magnetosphere. Nevertheless, because the rotation rate on the star varies with colatitude,
the Goldreich-Julian charge distribution in the magnetosphere differs slightly 
from the case of solid body rotation, as well as the
DC Poynting flux (Eq.(\ref{PoyntingDCgeneral})). Using as a model
the following even differential rotation:
\begin{equation}
v_\phi = R \sin \theta \left(\Omega_* + \delta \Omega_*  \sin \theta
\cos \Omega_T t\right)\, ,
\label{vphieven}
\end{equation}
we calculate $\Phi_{DC}$ from Eq. (\ref{PoyntingDCgeneral}) and integrate over the classical polar caps
to derive the emitted power ${\cal{P}}_{even}$:
\begin{equation}
{\cal{P}}_{even} = {\cal{P}}_{sol}
\left( 1 + \varepsilon_1 \cos \Omega_Tt + \varepsilon_2 \cos^2 \Omega_Tt\right) \, ,
\end{equation}
where
\begin{equation}
\varepsilon_1 = {8\over 5} \sqrt{\frac{\Omega_*R}{c}} 
\left(\frac{\delta \Omega_*}{\Omega_*}\right) \ ,
\quad
\varepsilon_2 = {2 \over 3}  \left(\frac{\Omega_*R}{c}\right)
\left(\frac{\delta \Omega_*}{\Omega_*}\right)^2 \ \cdot
\end{equation}
The wind power is modulated slightly at a level of $\varepsilon_1$, which is about 10 \%
for a period of 3 ms and $\delta \Omega_*/\Omega_* = \sqrt{\alpha_D} \approx 1/3$.
The change in the time-averaged power is at the 0.1 \% level for the same figures.
These small changes cannot account for the existence of a gamma-ray burst.

\subsection{Magnetosphere opening by an odd oscillation}
\label{opening}

An odd oscillation differs from an even one in that the two footpoints of a closed field line
experience differential motion in longitude, introducing a twist in this field line. This causes
a poloidal current to flow in the closed magnetosphere and drastically changes its
structure. When the twist exceeds a threshold of order $\pi$,
the magnetosphere opens. Field expansion by the shearing of 
the footpoints of field lines was
first discussed in the context of solar flares \citep{Hflares,Alyflares,Low1990}. It was
established that it occurs in Cartesian geometry with a direction of invariance
by theorems constraining the properties of line-tied force-free fields \citep{Alyflares,Aly1990}
and by numerical simulations \citep{BiskampWelter}.
The same process has been considered also in the case of
axisymmetric structures extending above a spherical surface on which the field lines are tied.
In general, it was demonstrated that rapid expansion occurs when a finite
shear is reached \citep{Alyaxisymm}. This is also supported by numerical
simulations \citep{MikicLinker}. Full opening occurs for a finite twist 
(of order $\pi$) in some specific examples \citep{LBBoily,Wolfson}.
In the present context, the light-cylinder radius imposes a limit on the distance to the apex of
closed field lines, such that field opening is even easier when magnetospheric inflation proceeds.
When the field opening becomes almost complete, it causes a growth in the polar caps and
the emitted wind power. In this case, the role of the torsional oscillation is
not to modulate the energy output by the addition of its own electromagnetic emission
but to open the door for a more significant wind emission from the central object.
This enhanced wind emission would acquire its energy directly from
the rotational  kinetic energy of the star, not only from the energy of the differential rotation,
and lasts for as long as the torsional oscillation survives with sufficient amplitude.

The magnetosphere  opens if the difference in longitude $\psi$
between the two conjugate footpoints of a field line exceeds
typically half a turn ($\psi >  \pi$). 
To justify this statement,
one should try to solve for the structure of the magnetosphere as a function of the
difference in longitude between the footpoints of field lines.
By assuming axisymmetry, the poloidal
field is represented by a flux function $a(r,\theta)$, 
so that the total magnetic field can be written as:
\begin{equation}
\vec{B} = \frac{\vec{\nabla}a \land \vec{e}_\phi}{r \sin \theta} 
+ B_\phi(r,\theta) \, \vec{e}_\phi \ \cdot
\label{Bdea}
\end{equation}
Any field line follows a surface of constant $a$ (a magnetic surface)
because the magnetic flux being transmitted through a circle perpendicular 
to and centred on the polar axis passing
at $(r,\theta)$, is $\Phi_M = 2 \pi a(r,\theta)$.
The perturbations of the closed magnetosphere of the star 
are of quasi-static nature (Sect. \ref{sectevenmodu}).
Neglecting the particle's inertia, the force equation
for the instantaneous equilibrium is the force-free equation:
\begin{equation}
c \, \rho_e \vec{E} +  \vec{j} \land \vec{B} = 0 \ \cdot
\label{eqforcefreemagnetosph}
\end{equation}
The electromagnetic state of the magnetosphere is not described by the
magnetic field alone but also by the electric potential $U(r, \theta)$.
Equation~(\ref{eqforcefreemagnetosph}) is supplemented by the time-independent
Maxwell's equations:
\begin{eqnarray}
\vec{\nabla} \land \vec{B} &=& 4\pi \vec{j}/c \ ,
\label{AmpereDC}
\\
\vec{\nabla} \cdot \vec{E} &=& 4 \pi \rho_e \ ,
\label{Poisson}
\\
\vec{E} = &-& \vec{\nabla} U \ .
\label{Potelec}
\end{eqnarray}
The components of Eq.~(\ref{eqforcefreemagnetosph})
can be expressed in terms of the functions $a(r, \theta)$, $U(r, \theta)$, and $I(r, \theta)$,
the latter being defined by:
\begin{equation}
I(r, \theta) = c \, r\sin \theta \, B_\phi \ \cdot
\label{definitionI}
\end{equation}
The current through the circle perpendicular to and centred on the polar axis
passing at $(r,\theta)$ is $J = I/2$. We refer to $I$ as
the poloidal current.
In an axisymmetric state, the electric field is poloidal. The toroidal component
of Eq.~(\ref{eqforcefreemagnetosph}) then shows that the gradients of $I$ and $a$
are everywhere parallel, which implies that $I$ is 
a function of $a$: $I(r, \theta) = I(a(r, \theta))$.
The poloidal part of Eq.~(\ref{eqforcefreemagnetosph}) can then be written as:
\begin{equation}
r^2 \sin^2 \theta \ \Delta U \ \, \vec{\nabla} U =
\left(\Delta^*a
+ \frac{I}{c^2} \, \frac{dI}{da} \right) \, \vec{\nabla} a \ ,
\label{eqpulsarspoloidale}
\end{equation}
where $\Delta$ is the scalar Laplacian and
\begin{equation}
\Delta^* a =
\frac{\partial^2 a}{\partial r^2} + \frac{\sin \theta}{r^2} \, \frac{\partial}{\partial \theta}
\left( \frac{1}{\sin \theta} \, \frac{\partial a}{\partial \theta} \right) \ \cdot
\end{equation}
Eq.~(\ref{eqpulsarspoloidale}) indicates that the gradients of $U$ and $a$
are everywhere parallel, which implies that $U$ is a function of $a$, 
$U(r, \theta) = U(a(r, \theta))$.
The rotation rate of the matter, $\Omega(a)$, is given by the electric drift velocity
of particles and is found to be:
\begin{equation}
\Omega(a) = c \ U'(a) \, \cdot
\end{equation}
The projection of Eq.~(\ref{eqpulsarspoloidale}) onto $\vec{\nabla}a$
provides the so-called pulsar equation \citep{Michelbook}:
\begin{eqnarray}
\left( 1 - \frac{r^2 \Omega^2 \sin^2 \theta}{c^2}\right) \Delta a
  &-& \frac{r^2 \sin^2 \theta}{c^2} \, \Omega \, \Omega'
\nonumber \\
- \frac{2}{r} \, \frac{\partial a}{\partial r}
- \frac{2 \cos \theta}{r^2 \sin \theta} \frac{\partial a}{\partial \theta}
&+& \frac{I \, I'(a)}{c^2}  = 0 \ \cdot
\label{eqpulsarsgeneralisee}
\end{eqnarray}
This equation has a singularity at the light-cylinder, which
causes any field line reaching this limit to diverge \citep{grecs}.
To determine approximately which field lines become open,
it suffices to solve Eq.~(\ref{eqpulsarsgeneralisee}) for  $r \Omega \sin \theta \ll c$
and confirm which field lines reach a distance larger than $c/\Omega_*$.
In this limit, Eq.~(\ref{eqpulsarsgeneralisee}) reduces to:
\begin{equation}
\frac{\partial^2 a}{\partial r^2} + \frac{\sin \theta}{r^2} \, \frac{\partial}{\partial \theta}
\left( \frac{1}{\sin \theta} \, 
\frac{\partial a}{\partial \theta} \right) + \frac{I \, I'(a)}{c^2} = 0 \ \cdot
\label{GradShaf}
\end{equation}
It is shown in Appendix \ref{appendixopening} that,
for a self-similar model of the magnetospheric field,
there is no solution to Eq.(\ref{GradShaf}) with closed field lines
when the twist exceeds $\pi$.

We represent odd differential rotation by the following simple 
model for surface differential rotation:
\begin{equation}
\delta \Omega(\theta) =  \delta \Omega_* \, \sin 2\theta \cos \Omega_T t \ \cdot
\label{modelerotdiff}
\end{equation}
The magnetic flux distribution on the surface of the star
is a function of the colatitude $\theta$, so that $\theta$ is a known function of $a$, $\theta(a)$.
The twist at time $t$ associated with Eq. (\ref{modelerotdiff}) is:
\begin{equation}
\psi(a) =   \frac{2 \delta \Omega_*}{\Omega_T}\   \sin 2\theta(a)  \, \sin \Omega_Tt \ \cdot
\label{dphimodel}
\end{equation}
It is implied here that the footpoint $P_2$ is at the colatitude $\theta$ and 
that $P_1$ and $P_2$ are at the same longitude at $t=0$.
In the model presented in Appendix \ref{appendixopening}, 
we regard  $<\psi>$  at time $t$ as being
half the maximum twist implied by Eq. (\ref{dphimodel}).
The limit value of $\pi$ for the twist is reached for a rather small amplitude of the torsional
oscillation because the period of the latter is long compared to the spin period.
Similarly to Eq.~(\ref{amplitudealphaD}), we  parametrize $\delta \Omega_*$ as
$\delta \Omega_* = \sqrt{\alpha_{odd}} \ \Omega^*$ in terms of the
fraction $\alpha_{odd}$ of the star's rotational energy 
available in this odd oscillation mode.
The magnetosphere is in an open magnetic configuration when:
\begin{equation}
\sin \Omega_Tt > {\mathrm{10}}^{-2} \
\left(\frac{P_*}{ {\mathrm{10}}^{-3} \, {\mathrm{sec}} }\right)\,
\left(\frac{10\, {\mathrm{sec}}}{P_T}\right) 
\left( \frac{{\mathrm{10}}^{-4}}{\alpha_{odd}}\right)^{1/2}
\ \cdot
\label{openingcriterium}
\end{equation}
As explained in Sect. \ref{sectevenmodu}, a 
fraction 10$^{-4}$ of the star's rotational energy may be stored
in odd torsional oscillation modes.
At this level of excitation of odd modes,
the magnetosphere would be open during a large fraction of the oscillation period.
There would be no opening only when $\alpha_{odd}$ is very small, i.e.:
\begin{equation}
\alpha_{odd} < {\mathrm{10}}^{-8} \
\left(\frac{P_*}{ {\mathrm{10}}^{-3} \, {\mathrm{sec}} }\right)^2
\left(\frac{10\, {\mathrm{sec}}}{P_T}\right)^2\, \cdot
\label{extinctionodd}
\end{equation}
We assume that odd oscillation modes are initially excited to a level
higher than the limit indicated by Eq. (\ref{extinctionodd}). 

A similar result is obtained when an even oscillation is considered
(with an amplitude given by the larger value indicated in Eq.
(\ref{amplitudealphaD})) but the dipolar-like magnetic-field lines are not strictly 
symmetric with respect to the equator.
This would happen if, for example, the field is a non-centred dipole.
We define $\delta \theta$ to be the difference of the 
absolute values of the latitudes of two conjugate footpoints.
The twist experienced by these footpoints 
will be larger than $\pi$, and thus the magnetosphere will open, 
when 
$\sqrt{\alpha_D} \, \Omega_* \ \left(\delta \theta/(\pi/2)\right) \, P_T > \pi$,
which translates into the condition: 
\begin{equation}
\frac{\delta \theta}{\pi/2} > {\mathrm{1.5}} \, \times  \, 10^{-4} 
\left(\frac{P_*}{ {\mathrm{10}}^{-3} \, {\mathrm{sec}} }\right)\,
\left(\frac{10\, {\mathrm{sec}}}{P_T}\right) \left(\frac{{\mathrm{10}}^{-1}}{\alpha_{D}}\right)^{1/2}
\ \cdot
\label{openingcritdecentered}
\end{equation}
For Eq. (\ref{openingcritdecentered}) to be  satisfied 
for typical values of $P_*$, $P_T$, and $\alpha_D$, 
it suffices that the dipole field
be decentred by a fraction of a few 10$^{-3}$ of the stellar radius.
Since the odd torsional oscillation has an amplitude larger than indicated by Eq. (\ref{extinctionodd})
or the magnetic 
geometry is north-south 
asymmetric to a degree larger than indicated by Eq. (\ref{openingcritdecentered}),
the magnetosphere will alternate during the oscillation cycle
between a classical state, which we refer to as closed, and an open state.

For an odd torsional oscillation, the configuration is closed when 
the condition of Eq. (\ref{openingcriterium}) is not satisfied. 
It becomes an open state, where all field lines are open and carry winds, when
the twist is sufficiently large for Eq. (\ref{openingcriterium}) to be satisfied.
During closed episodes, the polar caps opening is limited to $\theta_{pc} = (\Omega_* R/c)^{1/2}$,
and during open episodes $\theta_{pc} = \pi/2$. There is a
transitory state which we neglect because
it lasts much less than a wave period.
In an open state,
the power fed by the compact star into its relativistic wind
is much larger than the classical value given by Eq. (\ref{PoyntDC}).
This may be the reason why the emitted power is enhanced considerably
in the first moments after the collapse, an enhancement that should decline
as the star's rotation decelerates  and last at most
until the odd mode amplitude has decreased below the limit fixed by Eq. (\ref{extinctionodd}).
We calculate the lifetime of odd oscillations and their associated emission 
in Sect.\ref{dampingodd}.
The idea that a GRB would be the result of pulsar-type emission from a compact star with
an entirely open magnetosphere was considered by \citet{RudermanTK2000},
who however regard the expansion of the magnetosphere as being caused by
magnetic buoyancy rather than by twisting, as we suggest in this paper.

The power emitted at time $t$ is calculated by integrating the Poynting flux 
given in Eq. (\ref{PoyntingDCgeneral})
over the wind-emitting star surface, where $v_\phi$ is given by (see Eq.(\ref{modelerotdiff})):
\begin{equation}
v_\phi = R \Omega_* \sin \theta \,  \left(\Omega_* + \delta \Omega_*  \sin 2\theta \cos \Omega_T t\right) \ \cdot
\label{vsurfaceodd}
\end{equation}
When the magnetoshere is in a closed state, the emitted power is, neglecting terms
of order $(\delta \Omega_*/\Omega_*)^2$:
\begin{equation}
{\cal{P}}_{odd/cl}=  \frac{B_p^2 R^6 \Omega_*^4}{4 c^3} \ \cdot
\label{Lclosed}
\end{equation}
Similarly, when the magnetosphere is open, it is given by:
\begin{equation}
{\cal{P}}_{odd/op} =  \frac{2}{15} \ \frac{B_p^2 R^4 \Omega_*^2}{c} \ \cdot
\label{Lopen}
\end{equation}
Numerically, with $B_{p14}= B_p/$10$^{14}$ gauss and $R =$ 10 km:
\begin{equation}
{\cal{P}}_{odd/op} = {\mathrm{1.8 \ 10}}^{48} \ B_{p14}^2
\, \left( \frac{{\mathrm{10}}^{-3} \, {\mathrm{s}} }{P_*}\right)^2  \
{\mathrm{erg \, s}}^{-1} \ \cdot
\end{equation}
The power emitted during the open episodes is larger than 
that emitted during the closed episodes by a factor of $8 c^2/(15 \Omega_*^2 R^2)$.
For $P_*=$ 3 ms,
this factor is about 110. For the same rotation period and $ B_{p14} =$ 3,
${\cal{P}}_{odd/op}$ reaches 1.8 $\times$ 10$^{48}$ erg s$^{-1}$,
which is enough to explain the GRB emission, allowing for a
conversion factor from kinetic wind energy to
radiation that is smaller than unity.

\subsection{Damping of rotation and odd oscillation}
\label{dampingodd}

Neither the amplitude of the odd torsional oscillation
nor the rapid stellar rotation will last long in the presence of such large losses.
A rapidly spinning star with an odd oscillation is characterized by two parameters,
the average star-rotation rate $\Omega_*$ and the amplitude of the
differential rotation $\delta\Omega_*$. Due to wind losses, both decrease in time.
To calculate their evolution, a model of the internal magnetic field of the star is required. 
Although this is not an accurate model
for the winding-up of the field when the rotation depends only on the distance
to the axis, we shall assume for simplicity
that the unperturbed, magnetic field is uniform in the star and parallel to the
rotation axis, i.e. that $\vec{B}_< = B_p \, \vec{e}_z$. The normal component of this field 
on the star's surface equals that of a dipolar field and the inner field
$B_p$ equals the outer field at the $z>0$ pole. 
It is convenient for us to use the cylindrical coordinates 
$D$, $\phi$, and $z$, the parameter $r$ representing the spherical distance to the star's centre,
$R$ the radius of the star, and $\theta$ the colatitude. 
Since the fluid
is almost incompressible with a uniform mass density $\rho$,
we assume as in Sect. \ref{torsionaloscill} that its velocity 
is azimuthal and can be written as: 
\begin{equation}
\vec{V}  = (\Omega_* D + v_\phi(D, z, t))\ \, \vec{e}_\phi \, \cdot
\label{vitessephioddosc}
\end{equation}
The velocity $v_\phi$ is supposedly odd in $z$.  When writing the even
part of the rotation velocity as $\Omega_* D$, we 
neglect the even torsional modes, which play no role in the magnetosphere opening
in the case of a north-south, symmetrical, magnetic structure.
The magnetic field in the presence of the perturbation develops an azimuthal 
component equal to: 
\begin{equation}
\vec{B} = B_p \  \vec{e}_z + B_\phi(D, z, t) \ \vec{e}_\phi \, \cdot
\end{equation}
For the assumed, uniform, unperturbed field, 
Eqs.~(\ref{evolVphi})--(\ref{evolBphi}) can be written as:
\begin{eqnarray}
\frac{\partial v_\phi}{\partial t} &=& 
\frac{B_p}{4 \pi \rho} \, \frac{\partial B_\phi}{\partial z} \, ,
\label{AlfvenV}\\
\frac{\partial B_\phi}{\partial t} &=& B_p \, \frac{\partial v_\phi}{\partial z} \, \cdot
\label{AlfvenB}
\end{eqnarray}
If $v_\phi$ were, at any given time, structured according to the Proudman-Taylor theorem, it would
be cylindrical, that is, it would 
depend only on the distance to the rotation axis. 
Then, for a completely  axial field as assumed above, the term on the 
right-hand side of Eq. (\ref{AlfvenB}) would vanish.
The  assumption of a uniform axial field, however,
is only applied to simplify the following calculations.
Nothing constrains the structure of the field inside the star, especially when
it is not dynamically significant. For non-axial {\bf{B}}$_P$, Eqs.
(\ref{AlfvenV})--(\ref{AlfvenB}) 
still provide a sketchy representation of the torsional oscillation:
the operator $B_p \, \partial\!/\!\partial z$ must  be assumed to represent 
({\bf{B}}$_P\! \cdot\! \nabla$)
(see Eqs. (\ref{evolVphi})--(\ref{evolBphi})) and the variations in $r\sin \theta$
along a field line should be ignored, although they exist.
Using Eqs. (\ref{AlfvenV})--(\ref{AlfvenB}) is equivalent here to replacing
a curved poloidal magnetic field
in a cylindrical velocity
field with a uniform axial field in a $z$-dependent velocity field. 
An odd torsional oscillation in a cylindrical velocity field would correspond to
north-south asymmetric poloidal field lines.
The simple approach implied by Eqs. (\ref{AlfvenV})--(\ref{AlfvenB}) should be sufficient
for our purposes of estimating the damping time of the torsional oscillation.
These equations 
hold for a linear as well as for a non-linear axisymmetric perturbation
of an incompressible medium.
The velocity perturbation $v_\phi$ satisfies the Alfv\'en propagation equation:
\begin{equation}
\frac{\partial^2 v_\phi}{\partial t^2} = \frac{B_p^2}{4 \pi \rho} \, \frac{\partial^2 v_\phi}{\partial z^2} \, \cdot
\label{Alfvpropag}
\end{equation}
A standing wave solution of Eq.~(\ref{Alfvpropag}) is
\begin{equation}
v_\phi(D, z, t) = {\hat{v}}_\phi(D) \, \sin kz \cos \Omega_T t \, ,
\label{vphiAlf}
\end{equation}
where $k$ and $\Omega_T$ are related by
the dispersion relation $\Omega_T^2 =  k^2 v_A^2$
and $v_A^2= B_p^2/(4 \pi \rho)$.
The wavenumber $k$ could depend on the distance $D$ to the axis, 
each cylindrical magnetic surface then having its own oscillation period. 
To keep things simple, we assume that $k$ is a constant equal to
$\pi/R$, $\Omega_T = \pi v_A/R$,
and $v_\phi$ is an odd function of $z$. 
The proper choice of ${\hat{v}}_\phi(D)$ in Eq. (\ref{vphiAlf}) ensures that
the velocity field of the perturbation at the star's surface 
coincides with Eq.~(\ref{vsurfaceodd}), that is: 
\begin{equation}
{\hat{v}}_\phi(R \sin \theta) = \frac{2 R \, \delta \Omega_* \, 
\sin^2 \theta \cos \theta}{\sin (\pi \cos \theta)} \, \cdot
\label{hatvphi}
\end{equation}
Equations (\ref{AlfvenB}) and (\ref{vphiAlf}) infer the magnetic perturbation to be:
\begin{equation}
B_\phi = B_p \, \frac{{\hat{v}}_\phi}{v_A} \ \cos kz \, \sin \Omega_T t \, \cdot
\label{Bphioddosc}
\end{equation}
To obtain the angular-momentum balance equation for both the $z>0$ and
$z<0$ hemispheres, the torques acting on each must be calculated. 
Each hemisphere experiences volume torques exerted by magnetic tension 
and surface torques caused by the drag produced by
the emission of Poynting energy at the star's surface. 
The angular momentum $J_+$ of the $z>0$ hemisphere 
and the angular momentum $J_-$ of the $z< 0$ hemisphere can be calculated from 
Eqs. (\ref{vitessephioddosc}), (\ref{vphiAlf}), and (\ref{hatvphi}). They are:
\begin{eqnarray}
J_+ &=& I_1 \, \Omega_* + I_2 \ \delta \Omega_* \cos \Omega_T t \, ,
\\
J_- &=& I_1 \, \Omega_* - I_2 \ \delta \Omega_* \cos \Omega_T t \, \cdot
\end{eqnarray}
The moments of inertia $I_1$ and $I_2$ 
are:
\begin{eqnarray}
I_1 &=& \frac{4 \pi}{15} \, \rho R^5 \, ,
\label{defI1} \\
I_2 &=& 4 \, \rho R^5 \ \int_0^{\pi/2}\!\!  \sin^4\! \theta \, \cos^2\! \theta \ 
\tan\left(\frac{\pi}{2}\cos\! \theta\right) \ \, d \theta \, \cdot
\label{defI2}
\end{eqnarray}
The numerical value of the integral on the right-hand side of Eq.~(\ref{defI2}) is 0.187. 
The moment of inertia $I_*$ of the entire star with respect to the rotation axis is $I_* = 2 I_1$. 
The torque $T_{B+}$ exerted by magnetic tension 
on the $z>0$ hemisphere can be calculated from the Lorentz
force density (Eq. (\ref{AlfvenV})). If
$B_\phi$ is given by Eq. (\ref{Bphioddosc}), 
this implies that:
\begin{equation}
T_{B+} \! = \!- \, \frac{B_p^2 R^4}{v_A} \, \delta \Omega_* \sin\! \Omega_T t 
\! \int_0^{\pi/2}\!\! 
\sin^4\! \theta \cos^2\! \theta \tan\left( \frac{\pi}{2}\cos \theta \right)  d \theta  \cdot
\end{equation}
The magnetic tension torque $T_{B-}$ exerted on the $z < 0$ hemisphere is $ T_{B-} = - T_{B+}$.
The Poynting torque $dT_{P+}$ exerted on the strip of the star's surface between colatitudes $\theta$ and $\theta + d \theta$
in the $z>0$ hemisphere is related to the Poynting power $d{\cal{P}}_{+}$ emanating from that strip by
$d{\cal{P}}_{+} \!= \! - \, {\dot{\phi}} \, dT_{P+}$, 
where ${\dot{\phi}}$ is the angular velocity of the fluid
at that colatitude and at that time (see Eqs.~(\ref{vsurfaceodd}), (\ref{PoyntingDCgeneral}) and 
(\ref{vphiAlf})--(\ref{hatvphi})).
The torque $dT_{P-}$ on the strip between $\theta$ and $\theta + d \theta$
in the $z < 0$ hemisphere is similarly calculated:
\begin{eqnarray}
dT_{P+} \! &=&  - \, \frac{B_p^2 R^4}{2c} ( \Omega_* + \delta \Omega_* \sin 2 \theta \cos \Omega_T t) 
\sin^3\! \theta \cos^2\! \theta \, d\theta \, ,
\label{dTP+} \\
dT_{P-} \! &=& - \, \frac{B_p^2 R^4}{2c} ( \Omega_* - \delta \Omega_* \sin 2 \theta \cos \Omega_T t)
\sin^3\! \theta \cos^2\!  \theta \, d\theta \, \cdot
\label{dTP-}
\end{eqnarray}
The total Poynting torques $T_{P+}$ and $T_{P-}$ 
on the $z>0$ and $z<0$ hemispheres are derived by integrating 
Eqs. (\ref{dTP+}) and (\ref{dTP-}), respectively, over 
colatitudes from zero to the polar cap angle $\theta_{pc}$. 
When the magnetosphere is closed, 
$\theta_{pc} \approx (\Omega_* R/c)^{1/2}$.
When it is entirely open, $\theta_{pc} = \pi/2$. 
The angular-momentum balance equation for the $z>0$ and $z<0$ hemisphere are given respectively by:
\begin{eqnarray}
\frac{dJ_+}{dt} &=& T_{B+} + T_{P+} \, , \label{eqangmom+}\\
\frac{dJ_-}{dt} &=& T_{B-} + T_{P-} \, \cdot \label{eqangmom-}
\end{eqnarray}
By adding the expressions in Eqs. (\ref{eqangmom+}) and (\ref{eqangmom-}),
we derive an equation for the time evolution of the global rotation $\Omega_*$: 
\begin{equation}
\frac{d \Omega_*}{dt} = - \, \left(\frac{B_p^2 R^4}{I_* c} \ 
\int_0^{\theta_{pc}}\!\! \sin^3\! \theta \, \cos^2\! \theta \, d\theta
\right) \ \Omega_* \, \cdot
\label{eqOmega*}
\end{equation}
By substracting Eq. (\ref{eqangmom-}) from Eq. (\ref{eqangmom+}) we derive
after some algebra:
\begin{equation}
\frac{d\delta \Omega_*}{dt} = -  \,  \left(\frac{B_p^2 R^4}{I_2 c} \ 
\int_0^{\theta_{pc}}\!\!  \sin^4\! \theta \, \cos^3 \!\theta \, d\theta
\right) 
\ \ \delta \Omega_* \, \cdot
\label{eqdeltaOmega*}
\end{equation}
A characteristic damping time $\tau$ appears to be defined by:
\begin{equation}
\tau = \frac{cI_*}{B_p^2 R^4} = 
3\, \times \, 10^{3} \ \,  I_{45} B_{p14}^{-2} R_{10}^{-4} \, {\mathrm{sec}}\cdot
\label{taudampnum}
\end{equation}
When the magnetosphere is completely open,
Eqs.~(\ref{eqOmega*})--(\ref{eqdeltaOmega*}) reduce to:
\begin{eqnarray}
\frac{d \Omega_*}{dt} &=& - \, \frac{2}{15 \tau} \ \Omega_* \, ,
\label{evolOmegaopen} \\
\frac{d\delta \Omega_*}{dt} &=& -  \,  \frac{2 I_*}{35\, I_2 \, \tau} 
\ \delta \Omega_* \, \cdot
\label{evoldeltaOmegaopen}
\end{eqnarray}
From Eqs.~(\ref{defI1})--(\ref{defI2}), we find that the 
damping times for $\Omega_*$ and $\delta \Omega_*$ in the open regime are, respectively:
\begin{eqnarray}
\tau_{brake} &=& \frac{15 \tau}{2} = {\mathrm{7.5}} \, \tau \, ,
\label{tbrake}\\
\tau_{odd}&=& \frac{15 \tau}{2 \pi} \ 
\frac{\int_0^{\pi/2}\! \sin^4\! \theta \cos^2\! \theta \, \tan\left(\frac{\pi}{2}\cos \theta\right) \, d\theta
}{
\int_0^{\pi/2}\! \sin^4\! \theta \cos^3\! \theta \,d\theta} = {\mathrm{7.8}} \, \tau \, \cdot
\label{todd}
\end{eqnarray}
When the fast-spinning aligned rotator experiences episodes of magnetospheric opening,
its evolution  consists of a succession of
open (or high) states and closed, classical (low) states.
During open periods, Eqs.~(\ref{evolOmegaopen})--(\ref{evoldeltaOmegaopen}) apply
and the energy output, of the order indicated by Eq.~(\ref{Lopen}),
is considerable. During these periods, the open field should occasionally reconnect,
attempting to return to a closed structure, but, once reformed, the latter is again blown open
after the very short time needed to build a twist again 
of approximately half a turn. Large irregular variability
is then expected during these open periods, down to the millisecond timescale, which is the time
to cross through a light-cylinder size, expected 
to be representative of the equatorial current sheet, at the speed of light.
We note that the closed episodes are initially short in duration when
$\alpha_{odd}$ is of the order of 10$^{-4}$ as supected.
When a total time of order of $\tau_{odd}$ has been spent in the open state, the
oscillation has damped to an amplititude insufficient to open the magnetosphere, and
the average rotation has been substantially reduced. Neglecting the weak damping experienced
during the closed episodes, the spin rate $\Omega_*(t)$ and the largest amplitude
of the differential rotation $\delta \Omega_*(t)$ vary as
\begin{eqnarray}
\Omega_*(t) &=& \Omega_{*0} \ \ e^{-t/\tau_{brake}} \, ,
\label{despindet}
\\
\delta \Omega_*(t) &=& \delta \Omega_{*0} \ e^{- t/\tau_{odd}} \, \cdot
\end{eqnarray}
where $t$ is the cumulated time spent in the open state.
The emitted power scales as $\Omega_*^2$, and declines in a time of approximately $15 \tau/4$.
The GRB would disappear from view in about this time, 
which, for a polar  field of 3 $\times$ 10$^{14}$ gauss, is about 20 minutes.
The openings completely cease when the maximum twist falls below half a turn,
that is, 
from Eq.(\ref{dphimodel}), 
when $\delta \Omega_*/\Omega_T \sim  \pi$~.
This happens after a time $t_{stop}$
such that
\begin{equation}
t_{stop} = \tau_{odd} \ \ln\left(\frac{\delta \Omega_{*0}}{\pi \, \Omega_T}\right) =
\tau_{odd} \ \ln\left(\frac{\sqrt{\alpha_{odd}} \, P_T}{\pi \, P_{*0}}\right) \ \cdot
\label{tstop}
\end{equation}
The mean rotation rate $\Omega_{*stop}$ at time $t_{stop}$ is given by Eq.~(\ref{despindet}).
We define $ \xi = \tau_{odd}/\tau_{brake}$, which, from Eqs.~(\ref{tbrake})--(\ref{todd}), is nearly unity. 
Relating $\delta \Omega_{*0}$ to $\alpha_{odd}$ as in
Eq.~(\ref{openingcriterium}), we obtain, for $\xi =$ 1, 
$P_{*stop}  \approx \sqrt{\alpha_{odd}}\, P_T/\pi$.
Numerically:
\begin{equation}
P_{*stop} \approx   30 \, {\mathrm{ms}} \ \,
\Big(\frac{\alpha_{odd}}{{\mathrm{10}}^{-4}}\Big)^{1/2}
\Big(\frac{P_T}{{\mathrm{10\, s}} }\Big) \, \cdot 
\end{equation}
For $\alpha_{odd} =$ 3 $\times$ 10$^{-4}$ and $P_T =\,$ 0.8 s (i.e. half the period 
(\ref{estimationPT}) for
$B_p =$ 3 10$^{14}$ gauss) 
$P_{*stop} \approx$  4 ms.
After spending a time  $t_{stop}$
in the open state, the high episodes cease and the object becomes a pulsar
with a period of the order of 5 milliseconds.
As indicated above, the time of high activity is about 3.5 $\tau$ (Eq.~(\ref{taudampnum})),
which is comparable to the observed timescale of long duration GRBs when the
field of the compact star is somewhat higher than 10$^{14}$ gauss. For high fields,
of order 10$^{15}$ gauss, this timescale is about 100 seconds.

\section{Conclusion}
\label{Conclusion}

It is natural to consider 
that a new-born quark star experiences differential rotation,
causing its internal wound-up toroidal field to increase in strength to about 10$^{16}$ gauss.
This motion then develops into a  magnetic torsional oscillation, which
could be the origin of long-duration $\gamma$-ray bursts.
We have indeed shown that
an odd oscillation of small amplitude,
which should be easily reached, is sufficient to open the star's magnetosphere. 
A similar effect would also result from other causes of north-south asymmetries.
The rapid rotation
then drives a relativistic wind from the entire stellar surface.
When the star is a quark star, this wind is entirely
leptonic. 
We have calculated the Poynting power released and the timescale 
of this phenomenon, which meet the observational constraints if
the polar field of the quark star is of the order of a few $10^{14}$ gauss and its initial
rotational angular velocity  is of the order of 300 Hz. Large amplitude variations in
the light curve on timescales ranging from minutes to milliseconds
is a natural outcome of this process.

\subsection*{Acknowledgements}
\label{acknowledgements}
We are very grateful to L.J. Zdunik for his help in clarifying 
the conditions under which magnetic buoyancy is  inhibited by the finite
relaxation time of weak reactions among quarks.
MB was partially supported by the LEA Astro-PF collaboration
and the Marie Curie Intra-European Fellowships MEIF-CT-2005-023644 and ERG-2007-224793
within the 6th and 7th European Community Framework Programmes. This
work was supported in part by the MNiSW grant N20300632/0450.

{}

%%Appendix

\begin{appendix}

\section{Magnetosphere opening: an example}
\label{appendixopening}

\noindent
We describe the asymptotic properties
of the solutions to Eq. (\ref{GradShaf})
in the context of a self-similar model and show 
that there is a limit twisting for closed solutions to exist.
We define $A$ to be the equatorial value of the flux function $a$ (Eq. (\ref{Bdea})).
When a field line on the magnetic surface $a$ is twisted, there is a relation between
its twist $\psi(a)$, which is
the difference in longitude between its footpoints,
and the poloidal current $I(a)$. 
The differential equation of a field line is indeed:
\begin{equation}
\frac{dr}{B_r} = \frac{r\, d\theta}{B_\theta} = \frac{r \sin \theta \, d\phi}{B_\phi} \, \cdot
\label{eqdiffligneB}
\end{equation}
Using Eqs.~(\ref{Bdea}) and (\ref{definitionI}), we evaluate the change
in longitude $\psi$ accumulated
following a field line on the magnetic surface $a$ from one of its footpoints $P_1$ to the other $P_2$:
\begin{equation}
\psi(a) = I(a)\,  \int_{P_1}^{P_2} \frac{d\ell_P}{r^2 \sin \theta \mid B_P\mid} \, ,
\label{decalage}
\end{equation}
where $d\ell_P$ is the line element along the
poloidal field line $a$. 
It is the twist $\psi$, not the poloidal current $I$, which is known.
The relation Eq. (\ref{decalage}) does not infer $I(a)$ directly when $\psi(a)$
is known, because the line of constant $a$
on which the integration in Eq. (\ref{decalage}) is to be carried is unknown before the problem
expressed in Eq. (\ref{GradShaf}) has been solved.
To determine under which conditions a dipolar field line closing
at a few stellar radii would be  inflated sufficiently by the magnetospheric current
to reach the light-cylinder, we attempt to identify
separable solutions to Eq.~(\ref{GradShaf}) \citep{LBBoily, Wolfson, BardouH} of the form:
\begin{equation}
a(r,\theta) = A \left( \frac{R}{r}\right)^p \, g(\theta) \, \cdot
\label{ansatz}
\end{equation}
\citet{Wolfson} numerically studied the solutions of Eq.~(\ref{GradShaf}) 
under the ansatz (\ref{ansatz}).
We show here that the solutions must open when the twist reaches a finite value.
The largest value of $a$ being the total star flux $A$, 
$g(\theta) \leq 1$, and $g(\pi/2) = 1$. Similarly $g(0) = 0$, since there is no flux through
a circle of zero radius centred on the polar axis.
The apex of field line $a$ is at a distance $D(a)$, such
that $D(a) = R \, (A/a)^{1/p}$~: the more inflated the magnetosphere, the smaller the parameter $p$.
We will then deal with the small $p$ limit.
A solution of the form of Eq. (\ref{ansatz}) cannot match
any given flux distribution on the star, nor any given twist $\psi(a)$.
The constraint of Eq.(\ref{decalage}) can only be satisfied in an average sense.
Using Eq. (\ref{ansatz}) in Eq. (\ref{GradShaf}) the following equation is obtained:
\begin{equation}
\frac{p (p + 1)}{g^{2/p}} + \frac{\sin \theta}{g^{1 + 2/p}} \frac{\partial}{\partial \theta}
\! \left(\! \frac{1}{\sin \theta}  \frac{\partial g}{\partial \theta} \!\right) =
-  \frac{A^{2/p}}{c^2/ R^2} \frac{I(a) I'(a)}{a^{1 + 2/p}} \, \cdot
\label{thetaegala}
\end{equation}
The left-hand side is a function of $\theta$, while the right-hand side is a function of $a$.
Both then equal a common constant, $-K$. 
This means that for the ansatz of Eq. (\ref{ansatz}) 
to be satisfied when the similarity exponent is $p$,
the current function $I(a)$ must be:
\begin{equation}
\frac{I(a)}{c} = \sqrt{\frac{Kp}{p + 1} } \ \frac{a^{1 + 1/p}}{A^{1/p} R} \, \cdot
\label{solselfsimI}
\end{equation}
The dipolar angular function $g= \sin^2 \theta$
is recovered for $p=1$ and $I = K = 0$. For non-vanishing $K$, the constraint 
of Eq. (\ref{decalage})
becomes, for solutions of the form Eq. (\ref{ansatz}):
\begin{equation}
\frac{I(a)}{c} =  \psi(a) \ 
\frac{a^{1 + 1/p}}{A^{1/p} R} \ \frac{p}{\int_0^\pi g^{1/p}(\theta) \, d\theta} \, \cdot
\label{twistIselfsim}
\end{equation}
The limits on the integral at the denominator reflects the fact
that all field lines span the interval $[0, \pi]$ in $\theta$ (Eq. (\ref{ansatz})).
Equation~(\ref{twistIselfsim}) is consistent with Eq. (\ref{solselfsimI}) only when $\psi(a)$ is
independent of $a$, which corresponds to the peculiar twist profile 
in which one hemisphere rotates like a solid body and
the other in an opposite sense. We define $<\psi>$ to be this twist. We may think of 
$<\psi>$ as being some average on one hemisphere of a more realistic twist profile.
$K$ is related to
$<\psi>$ by:
\begin{equation}
\sqrt{\frac{Kp}{p + 1}} =  \frac{p \, <\psi>}{\int_{0}^{\pi} g^{1/p}(\theta) \, d\theta} \, \cdot
\label{Kettwistselfsim}
\end{equation}
Having chosen a value of $<\psi>$ this relation infers $K$ for a given $p$.
Given this relation, the value of $p$ itself results from
the need for the solution of the angular function $g(\theta)$ to
satisfy the three
requirements $g(\pi/2) = 1$, $g(0) = 0$, and $g'(\pi/2) = 0$, the last one
resulting from the symmetry of magnetic surfaces with respect to the equator.
A second-order differential equation accepting only two boundary conditions,
the extra condition eventually determines the value of $p$. To express this condition  explicitly,
Eq.~(\ref{thetaegala}) for the angular function $g$ must be solved in the limit
of small $p$. 
In terms of the variable $x = \cos \theta$, Eq. (\ref{thetaegala}) can be written:
\begin{equation}
(1 - x^2) \, \frac{d^2 g}{d x^2} + p(p + 1) g + K g^{1 + 2/p} = 0 \, \cdot
\label{eqgmu}
\end{equation}
For small $p$ the second term of Eq. (\ref{eqgmu}) is negligible.
The exponent $(1 + 2/p)$ being very large, the third term of (\ref{eqgmu})
essentially vanishes wherever $g < 1$. It remains non-negligible only in
the vicinity of the equator ($x =0$), where $g$ reaches unity.
The solution $g(x)$ is then almost  a linear function
for all $x$, except in a small region about $x = 0$. This allows us to
simplify Eq. (\ref{eqgmu}) in the small $p$ limit as:
\begin{equation}
\frac{d^2 g}{d x^2} + K g^{1 + 2/p} = 0\, \cdot
\label{eqgmupetitp}
\end{equation}
This equation has a first integral. The condition
that $g'(0) = 0$ at $x = 0$, where $g$ must equal unity, can be satisfied
by an appropriate choice of the integration constant, giving:
\begin{equation}
{g'}^2 = \frac{Kp}{p+1} \, \Big( 1 - g^{2 + 2/p} \Big) \, \cdot
\label{intpremiere}
\end{equation}
Wherever $g$ is sufficiently less than unity, ${g'}^2\approx Kp$. However,
we know that in these regions
the modulus of the slope should be unity, because $g$ has already been recognized to be
a  linear function varying from $g = 0$ at $x =1$ to very nearly $g=1$ at $x =0$. 
The relation between $K$ and $p$ is then, in the small $p$ limit:
\begin{equation}
Kp = 1 \, \cdot
\label{Kpegal1}
\end{equation}
We should now establish the relation between $<\psi>$ and $p$ resulting from
Eq.~(\ref{Kettwistselfsim}). To calculate the angular integral at
the denominator, we must solve Eq.~(\ref{intpremiere}) when $p \ll 1$.
Taking Eq. (\ref{Kpegal1}) into account, changing the unknown function $g$
for $h$ such that $g = (1- ph)$ and making use of the fact that
the limit for $p$ approaching 0 of $(1+ py)^{1/p}$ is $\exp(y)$, it can be shown that
the solution of Eq.~(\ref{intpremiere}) is in this limit:
\begin{equation}
g = 1 - p \ {\mathrm{ln}} \left( \cosh \left( \frac{x}{p}\right) \right) \, \cdot
\label{gfinalementpetitp}
\end{equation}
Since $p$ is small, $ g^{1/p} \approx 1/\cosh(x/p)$.
The integral that appears in Eq. (\ref{Kettwistselfsim})
can then be calculated, resulting in:
\begin{equation}
\lim_{p \rightarrow 0} \, < \psi > = \pi \, \cdot
\label{psiegalpi}
\end{equation}
Thus, the exponent $p$ approaches zero
as the twist approaches $\pi$ and the
magnetosphere swells boundlessly in this limit.
For our purpose, it is sufficient that
a field line extends farther than the light-cylinder for it to open.
We can then safely adopt the limit of a twist of a half a turn in causing an almost
complete opening.

\end{appendix}


\begin{thebibliography}{}

\bibitem[Aksenov et al.(2003)]{AksenovMU03}
Aksenov, A.G., Milgrom, M., Usov, V.V., Mon. Not. Roy. Astron. Soc. {\bf 343}, L69 (2003)

\bibitem[Alcock et al.(1986a)]{AlcockFO86}
Alcock, C., Farhi, E., Olinto, A.V., ApJ {\bf 310}, 261 (1986a)

\bibitem[Alcock et al.(1986b)]{AlcocketalPhysRevL}
Alcock, C., Farhi, E., Olinto, A.V. Phys. Rev. Letters {\bf{57}}, 2088 (1986b)

\bibitem[Alford et al.(2007)]{Alford}
Alford, M.G.,  Schmitt, A., Rajagopal, K., Schafer, T., to appear in 
Rev. Mod. Phys., arXiv:0709.4635 [hep.ph] (2007)

\bibitem[Aly(1985)]{Alyflares}
Aly, J.J., Astron. Astrophys. {\bf 143}, 19 (1985)

\bibitem[Aly(1990)]{Aly1990}
Aly, J.J., Comput.Phys.Comm., {\bf 59}, 13 (1990)

\bibitem[Aly(1994)]{AlyBopBpo}
Aly, J.J., Astron. Astrophys. {\bf 288}, 1012 (1994)

\bibitem[Aly(1995)]{Alyaxisymm}
Aly, J.J., ApJ {\bf 439}, L66 (1995)

\bibitem[Asseo et al.(1975)]{Estelle75}
Asseo, E.,  Kennel, F.C., Pellat, R., Astron. Astrophys. {\bf 44}, 31 (1975)

\bibitem[Bardou \& Heyvaerts(1996)]{BardouH}
Bardou, A., Heyvaerts, J., Astron. Astrophys. {\bf 307}, 1009 (1996)

\bibitem[Bastrukov \& Podgainy(1996)]{Bastrukov}
Bastrukov, S.I., Podgainy, D.V., Phys. Rev. E {\bf 54}, 4465 (1996)

\bibitem[Bejger \& Haensel(2002)]{MichalPawel}
Bejger, M., Haensel, P., Astron. Astrophys. {\bf 396}, 917 (2002)

\bibitem[Berezhiani et al.(2002)]{Berezhiani02}
Berezhiani, Z., Bombaci, I., Drago, A., Frontera, F.,
Lavagno,  A.,
Nucl. Phys. B - Proceedings Supplements {\bf 113}, 268 (2002)

\bibitem[Berezhiani et al.(2003)]{Berezhiani03}
Berezhiani, Z., Bombaci, I., Drago,  A., Frontera, F.,
Lavagno,  A., ApJ {\bf 568}, 1250 (2003)

\bibitem[Biskamp \& Welter(1989)]{BiskampWelter}
Biskamp, D., Welter, H., Solar Physics {\bf 120}, 49 (1989)

\bibitem[Bonazzola et al.(2007)]{BVB07}
Bonazzola, S., Villain, L., Bejger, M.,
Class. Quantum Grav. {\bf 24}, S221 (2007)

\bibitem[Bombaci \& Datta(2000)]{BombaciDatta00}
Bombaci,  I., Datta,  B.,
ApJ {\bf 530}, L69 (2000)

\bibitem[Bucciantini et al.(2006)]{buc06}
Bucciantini, N., Thompson, T.~A., Arons, J.,
Quataert, E., del Zanna, L., Mon.\ Not.\ Roy.\ Astron.\ Soc. {\bf 368}, 1717 (2006)

\bibitem[Burrows et al.(2007)]{Burrowsetal2007}
Burrows, A., Dessart, L., Livne, E., Ott, C. D., Murphy, J.,
ApJ {\bf 664}, 416 (2007)

\bibitem[Cheng \& Dai(1996)]{ChengDai96}
Cheng, K.S., Dai, Z.G., Phys. Rev. Lett. {\bf 77}, 1210 (1996)

\bibitem[Cheng \& Dai(1998a)]{ChengDai98a}
Cheng, K.S., Dai, Z.G., Phys. Rev. Lett. {\bf 80}, 18, (1998a)

\bibitem[Cheng \& Dai(1998b)]{ChengDai98b}
Cheng, K.S., Dai, Z.G., Phys. Rev. Lett. {\bf 81}, 4301 (1998b)

\bibitem[Contopoulos et al.(1999)]{grecs}
Contopoulos, I., Kazanas, D., Fendt, C., ApJ {\bf 511}, 351 (1999)

\bibitem[Dai \& Lu(1998a)]{DaiLu98}
Dai,  Z.G., Lu,  T., Phys. Rev. Lett. {\bf 81}, 261 (1998a)

\bibitem[Dai \& Lu(1998b)]{dl98}
Dai, Z.G., Lu, T., Phys. Rev. Lett. {\bf 81}, 4301 (1998b)

\bibitem[Dar \& De R\'ujula(2004)]{DarDeRujula04}
Dar, A., De R\'ujula, A., Physics Reports, {\bf 405} 203 (2004)

\bibitem[Dar(2006)]{Darreview}
Dar, A., Chinese J. Astron. Astrophys. {\bf 6}, Suppl.1, 301 (2006) 

\bibitem[De R\'ujula(1987)]{DeRujula87}
De R\'ujula, A., Phys. Lett., {\bf 193}, 514 (1987)

\bibitem[Dessart et al.(2007)]{dessartetal2007}
Dessart, L., Burrows, A., Livne, E., Ott, C. D.,
ApJ {\bf 669}, 585 (2007)

\bibitem[Drago et al.(2004a)]{DragoLP04a}
Drago, A., Lavagno, A., Pagliara, G.,
AIP Conf. Proceedings {\bf 727}, 420 (2004a)

\bibitem[Drago et al.(2004b)]{DragoLP04b}
Drago, A., Lavagno, A., Pagliara, G.,
Phys. Rev. D  {\bf 69}, 057505 (2004b)

\bibitem[Drago et al.(2006)]{DragoLP06}
Drago, A., Lavagno, A., Pagliara, G.,
Nucl. Phys. A {\bf 774}, 823 (2006)

\bibitem[Drago et al.(2007)]{DragoLP07}
Drago, A., Lavagno, A., Pagliara, G.,
Nucl. Phys. A {\bf 782}, 418 (2007)

\bibitem[Fryer \& Kalogera(2001)]{Fryer}
Fryer, C., Kalogera, V., ApJ {\bf 554}, 548 (2001)

\bibitem[Gaensler et al.(2005)]{Gaensler2005}
Gaensler, B.~M., McClure-Griffiths, M.~N., Oey, M.~S., Havekorn, M., Dickey, J.~M., Green, A.~G., 
ApJ {\bf 620} L95 (2005)

\bibitem[Goldreich \& Julian(1969)]{GoldJulian}
Goldreich, P., Julian, W.~H., ApJ {\bf157}, 869 (1969)

\bibitem[Haberl(2007)]{Haberl}
Haberl, F., Astrophys. \& Sp. Sc. {\bf 308}, 181 (2007)

\bibitem[Haensel et al.(1986)]{HZS86}
Haensel, P., Zdunik, J.L., Schaeffer, R., Astron. Astrophys. {\bf 160}, 121 (1986)

\bibitem[Haensel et al.(1991)]{HPA91}
Haensel, P., Paczy{\'n}ski, B., Amsterdamski, P., ApJ {\bf 375}, 209 (1991)

\bibitem[Haensel \& Zdunik(2007)]{hz07}
Haensel, P., Zdunik, J.L.,
Nuovo Cimento {\bf 121 B}, 1349 (2007)

\bibitem[Haensel et al.(2007)]{Haenseletal2007}
Haensel, P., Pothekin, A.Y., Yakovlev, D.G., Neutron stars 1. Equation of state and structure,
Springer Verlag, (2007)

\bibitem[Heiselberg \& Pethick(1993)]{Heiselberg}
Heiselberg, H., Pethick, C.J., Phys. Rev. D {\bf48}, 2916 (1993)

\bibitem[Heyvaerts et al.(1982)]{Hflares}
Heyvaerts, J., Lasry, J.M., Schatzman, M.,  Witomsky, P., Astron. Astrophys.\ {\bf 111}, 104 (1982)

\bibitem[Heyvaerts \& Norman(2003)]{HN2003}
Heyvaerts, J., Norman, C.A., ApJ {\bf 596}, 1240 (2003)

\bibitem[Iwamoto(1983)]{Iwamoto}
Iwamoto, N., Phys.Rev.D\ {\bf28}, 2353 (1983)

\bibitem[Klu\'zniak \& Ruderman(1998)]{kr98}
Klu\'zniak, W., Ruderman, M., ApJ {\bf 505}, L113 (1998)

\bibitem[Landau \& Lifshitz(1975)]{LandauLifchitz}
Landau, L.D. \&  Lifshitz, E.M., The classical theory of fields, Butterworth-Heinemann, (1975)

\bibitem[Low(1990)]{Low1990}
Low, B.C. ARA\&A, {\bf 28}, 205 (1990)

\bibitem[Lugones et al.(2002)]{Lugones2002}
Lugones,  G., Ghezzi,  C.R., de Gouveia dal Pino,  E.M.,
Horvath J.E., ApJ {\bf 581}, L101 (2002)

\bibitem[Lynden-Bell \& Boily(1994)]{LBBoily}
Lynden-Bell, D., Boily, C., Mon.\ Not.\ Roy.\ Astron.\ Soc.\ {\bf 267}, 146 (1994)

\bibitem[Madsen(1992)]{Madsen}
Madsen, J., Phys. Rev. D {\bf 46} 3290 (1992)

\bibitem[Madsen(2000)]{Madsen2000}
Madsen, J., Phys. Rev. Letters {\bf 85} 10 (2000)

\bibitem[M{\'e}sz{\'a}ros(2006)]{Mesz06}
M{\'e}sz{\'a}ros, P., Rep. Prog. Phys.  {\bf 69} 2259 (2006)

\bibitem[Michel(1969)]{Michel69}
Michel, F.C., ApJ {\bf 158}, 727 (1969)

\bibitem[Michel(1991)]{Michelbook}
Michel, F.C., The theory of Neutron Star Magnetospheres, U. of Chicago Press (1991)

\bibitem[Miki\'c \& Linker(1994)]{MikicLinker}
Miki\'c, Z., Linker, J.A., ApJ {\bf 430}, 898 (1994)

\bibitem[Obergaulinger et al.(2006a)]{Obergau2}
Obergaulinger, M., Aloy, M.A., M\"uller, E., Astron. Astrophys. {\bf 450}, 1107 (2006a) 

\bibitem[Obergaulinger et al.(2006b)]{Obergau1}
Obergaulinger, M., Aloy, M.A., Dimmelmeier, H., M\"uller, E., Astron. Astrophys. {\bf 457}, 209 (2006b)

\bibitem[Ouyed et al.(2002)]{OuyedDD02}
Ouyed, R., Dey, J., Dey, M.,
Astron. Astrophys. {\bf 390}, L39 (2002)

\bibitem[Ouyed \& Sannino(2002)]{OuyedSannino02}
Ouyed, R., Sannino, F.,
Astron. Astrophys. {\bf 387}, 725 (2002)

\bibitem[Paczy{\'n}ski(1990)]{PaczynskiSuperEdd}
Paczy{\'n}ski, B., ApJ {\bf 363}, 218 (1990) 

\bibitem[Paczy{\'n}ski \& Haensel(2005)]{PH2005}
Paczy{\'n}ski, B., Haensel, P., Mon.\ Not.\ Roy.\ Astron.\ Soc.\ Lett.\  {\bf 362}, L4 (2005)

\bibitem[Rajagopal \& Wilczek(2001)]{Rajagopal}
Rajagopal, K., Wilczek, F., Phys. Rev. Letters {\bf{86}}, 3492 (2001)

\bibitem[Rincon \& Rieutord(2003)]{Rieutord}
Rincon, F., Rieutord, M., Astron. Astrophys. {\bf 398}, 663 (2003)

\bibitem[Ruderman et al.(2000)]{RudermanTK2000}
Ruderman, M., Tao, L., Klu{\'z}niak, W., ApJ {\bf 542}, 243 (2000)

\bibitem[Salvati(1978)]{Salvati78}
Salvati, M., Astron. Astrophys. {\bf 65}, 1 (1978)

\bibitem[Shternin \& Yakovlev, 2006]{Shternin}
Shternin, P.S., Yakovlev, D.G., Phys. Rev. D {\bf 72}, 043004 (2006)

\bibitem[Spruit(1999)]{Spruit99}
Spruit, H.C., Astron. Astrophys. {\bf 341}, L1 (1999)

\bibitem[Steiner et al.(2001)]{Steineretal2001}
Steiner, A.W., Prakash, M., Lattimer, J.W., Phys. Lett. B  {\bf 509}, 10 (2001)

\bibitem[Thompson(1994)]{Thompson94}
Thompson, C., Mon.\ Not.\ Roy.\ Astron.\ Soc.\ {\bf 270}, 480 (1994)

\bibitem[Usov(1992)]{Usov92}
Usov, V.V., Nature {\bf 357}, 472 (1992)

\bibitem[Usov(2001)]{Usov01}
Usov, V.V., ApJ {\bf 550} L179 (2001)

\bibitem[Wang et al.(2000)]{WangDai00}
Wang,  X.Y., Dai,  Z.G., Lu,  T.,
Wei,  D.M., Huang,  Y.F.,
Astron. Astrophys. {\bf 357}, 543 (2000)

\bibitem[Wolfson(1995)]{Wolfson}
Wolfson, R., ApJ {\bf443}, 810 (1995)

\bibitem[Zhang \& M{\'e}sz{\'a}ros(2001)]{ZM01}
Zhang, B., M{\'e}sz{\'a}ros, P., ApJ {\bf 552}, L35 (2001)

\bibitem[Zhang \& M{\'e}sz{\'a}ros(2002)]{ZM02}
Zhang, B., M{\'e}sz{\'a}ros, P., ApJ {\bf 566}, 712 (2002)

\bibitem[Zhang \& M{\'e}sz{\'a}ros(2004)]{ZM04}
Zhang, B., M{\'e}sz{\'a}ros, P., Int. Jour. of Mod. Phys. A {\bf 19}, 2385 (2004)

\bibitem[Ziolkowski(2002)]{Ziolkowski}
Ziolkowski, J., Mem. Soc. Astron. Ital. {\bf 73}, 300 (2002)

\end{thebibliography}
\end{document}